\documentclass[a4paper,UKenglish,cleveref, autoref, thm-restate]{lipics-v2021}

\newif\ifLongVersion
\LongVersionfalse
\LongVersiontrue

\usepackage{amsmath,amssymb,amsthm}
\usepackage{stmaryrd}
\usepackage{proof}
\usepackage{tikz}
	\usetikzlibrary{calc}

\usepackage{FMC}

\newcommand{\commentout}[1]{}

\newcommand\looppic[1]{
\vcenter{\hbox{%
\begin{tikzpicture}[thick,x=20pt,y=12pt,outer sep=0pt,inner sep=2pt]
	\coordinate (m) at (0,0);
	
	\node at (-0.3,-2) {$\dots$};
	
	\begin{scope}[->,>=stealth,anchor=north]
		\draw (m)--node[pos=0.7,anchor=east]{$\termcolor\scriptstyle{i_1~}$} (-1,-2) node {$\vphantom:\typecolor\scriptstyle{\ifx#11p_1\else\frac{p_1}{1-q}\fi}$};
		\draw (m)--node[pos=0.7,anchor=east]{$\termcolor\scriptstyle{i_n\,}$} (.4,-2) node {$\vphantom:\typecolor\scriptstyle{\ifx#11p_n\else\frac{p_n}{1-q}\fi}$};
	\ifx#11
		\draw (m)--node[pos=0.7,anchor=west]{\,$\termcolor\scriptstyle{~j}$}   ( 1,-2) node {$\vphantom:\typecolor\scriptstyle{q}$};
	\fi
	\end{scope}
	
	\node (x) at (0,-3) {};
	
	\coordinate (a) at ( .85,-1.7);
	\coordinate (b) at (1.25,-1.6);
	
    	\ifx#10
		\draw[rounded corners=5pt] (m) -- (.5,-1) .. controls (a) and (b) .. (1.25,-.75) --node[pos=0.3,anchor=west]{\,$\termcolor\scriptstyle{j}$} (1.25,1.3) -- (0,1.3) -- (0,0.8);
    	\fi	
	
	\node[draw=red,fill=red!10,rounded corners,minimum size=16pt] (M) at (m) {$\term M$};
	\draw[->,>=stealth] (0,1.6)--(M);
\end{tikzpicture}}}
}

\newcommand\markov[1]{\llparenthesis\kern1pt\term{#1}\kern1pt\rrparenthesis}
\newcommand\sem[1]{\llbracket#1\rrbracket}

\newcommand\PFMC{\textrm{PFMC}}
\newcommand\PFMCT{\PFMC$^{\,\smallbin{\Rightarrow\oplus}}$}

\title{Probabilistic Simple Types for Almost-Sure Termination (Extended Version)}

\author{Ugo {Dal Lago}}{University of Bologna, Italy \and INRIA Sophia Antipolis, France}{ugo.dallago@unibo.it}{https://orcid.org/0000-0001-9200-070X}{}

\author{Willem Heijltjes}{University of Bath, United Kingdom \and \url{http://willem.heijltj.es}}{w.b.heijltjes@bath.ac.uk}{https://orcid.org/0009-0001-8941-1150}{}

\authorrunning{U. Dal Lago and W. Heijltjes}

\Copyright{Ugo Dal Lago and Willem Heijltjes}

\ccsdesc[500]{Theory of computation~Lambda calculus}
\ccsdesc[500]{Theory of computation~Probabilistic computation}

\keywords{probabilistic lambda-calculus, type systems, almost-sure termination}

\acknowledgements{We thank Georgina Majury for her efforts on a precursor to this work, and the anonymous referees for their comments, in particular those leading to the observations in Section 8.}

\nolinenumbers

\hideLIPIcs

\EventEditors{Corina C\^{\i}rstea and Yannick Forster}
\EventNoEds{2}
\EventLongTitle{35th EACSL Annual Conference on Computer Science Logic (CSL 2027)}
\EventShortTitle{CSL 2027}
\EventAcronym{CSL}
\EventYear{2027}
\EventDate{January 26--29, 2027}
\EventLocation{Brighton, UK}
\EventLogo{}
\SeriesVolume{403}
\ArticleNo{38}

\begin{document}

\maketitle

\begin{abstract}
We present a type system for almost-sure termination (AST) in higher-order probabilistic computation. Our vehicle is the recent Functional Machine Calculus (FMC), which combines the lambda-calculus with effectful imperative computation. Importantly for our purpose, it models branching and looping computation with sum types and a typed iterator construct. Our key observation is that by replacing sum types with probability distributions, types ensure that iteration has a non-zero chance of exiting, which then gives AST in a simple and natural way. 

Concretely, our type system assigns explicit probability distributions to termination outcomes, allowing probabilistic behaviour to be tracked compositionally at the type level. Types then soundly guarantee almost-sure termination, even in the presence of higher-order functions. The fragment of ground terms, which consists purely of the branching and iteration constructs with probabilistic sampling, is equivalent to finite Markov chains and is positively AST (expected evaluation time is finite). The full calculus is more expressive, which we demonstrate through an example term with infinite expected evaluation time.
\end{abstract}


\section{Introduction}

We are interested in applying the toolset of the $\lambda$-calculus, in particular confluent reduction and type systems, to probabilistic computation. In~\cite{DalLago-Guerrieri-Heijltjes-2020} we restored confluence to a probabilistic $\lambda$-calculus, the \emph{probabilistic event $\lambda$-calculus} (PEL), by decomposing probabilistic choice into a \emph{sampling} construct and a \emph{conditional}. Then in~\cite{Antonelli-DalLago-Pistone-2022} and~\cite{Heijltjes-Majury-2025}, from different inspirations we converged on essentially the same probabilistic type system for the PEL to give lower bounds on the probability of termination. In this paper we further extend the calculus, and develop a type system for \emph{almost-sure termination} (AST): termination with probability one. In doing so, our objective differs from that of many existing works in this area, which aim to capture as many AST programs as possible~\cite{DalLago-Grellois-2019,DalLago-Faggian-Ronchi-2021}, or even to study the feasibility of capturing all such programs~\cite{Kobayashi-DalLago-Grellois-2020}. Instead, we seek to understand the extent to which almost-sure termination can be obtained and explained in a natural manner, possibly using well-established methods, and guaranteed by a minimal extension to simple types. More powerful methods such as dependent types and intersection types may later build on this along standard lines.

We start from the following observation. \emph{Iteration}, generally represented by while-loops, repeats a given function $f$ until a condition is fulfilled. In the typed case, illustrated  below, the function $f$ is required to return a sum $B+A$, where the choice between returning a value $A$ or $B$ represents the condition. Then $\mathsf{iter}\,f$ evaluates $f$, exits on $B$, and repeats on $A$~\cite{Bloom-Esik-1993}.
\[
\begin{array}{@{}r@{}l@{}}
	f &{}: A \to B+A
\\ \hline
	\mathsf{iter}\,f &{}: A\to B
\end{array}
\]
With the above rule, as with typed fixed-point combinators, typing no longer guarantees termination. However, we observe that if we replace $B+A$ by a probabilistic sum $B \oplus_p A$ with a fixed probability $p\in (0,1]$, i.e.\ the probability of choosing $B$ is non-zero and does not depend on the input, then $\mathsf{iter}\,f$ is \emph{almost-surely terminating} (provided of course that $f$ is itself terminating). This opens a way to approach almost-sure termination via type systems. A crucial question arises: how can we handle a type constructor such as $B\oplus_p A$ within a $\lambda$-calculus with probabilistic choice in a compositional and elegant manner?

These considerations illustrate how probabilistic choice is connected to other imperative features such as sequentiality and iteration. Derived from the PEL, the second author has developed a new approach to unifying the $\lambda$-calculus with imperative computation and effects, the \emph{Functional Machine Calculus} (FMC)~\cite{Heijltjes-2023}. Its most recent instance~\cite{Heijltjes-2025-MFPS} features branching sequential computation and iteration, providing an ideal vehicle for our present study. In the FMC each branch of the computation is labelled with a \emph{choice} label $i,j,k,\dots$, which takes the r\^ole of a constant, an exception, a case label, or an exit status. For example, the boolean constants $\bot$ and $\top$ are choice labels. Iteration of a term $\term M$ is written $\term{M^i}$, which evaluates $\term M$ and repeats if $\term M$ chooses $i$, terminating for any other choice. The calculus encodes a fair probabilistic choice, which we write as $\term{M+N}$.

\begin{example}[Bernoulli Factory]
The term $\term{(\heads + (\tails + \,i))^i}$ chooses \emph{heads}, the choice $\heads$, with probability $\frac12$, \emph{tails} $\tails$ with probability $\frac14$, and iterates with probability $\frac14$, to give an overall probability of $\frac 23$ for $\heads$ and $\frac 13$ for $\tails$. We illustrate it as follows; it will remain a running example.
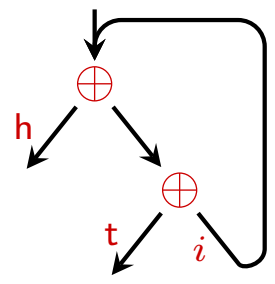
\[
	\term{(\heads + (\tails + \,i))^i}
\quad=\quad
\vc{%
\begin{tikzpicture}[thick,x=16pt,y=20pt,outer sep=0pt,inner sep=1pt,]
	\node (x) at (1,2) {$\termcolor\oplus$};
	\node (y) at (2,1) {$\termcolor\oplus$};

	\coordinate (a) at (2.75,0.25);
	\coordinate (b) at (3   ,0.3);
	
	\begin{scope}[->,>=stealth,anchor=north]
	
	\draw (1,2.7)--(x);
	\draw (x)--(y);
	\draw (x) --node[pos=0.6,anchor=south east]{$\termcolor\scriptstyle{\heads}\,$} (0.2,1.2);
	\draw (y) --node[pos=0.6,anchor=south east]{$\termcolor\scriptstyle{\tails}\,$} (1.2,0.2);
	
	\draw[rounded corners=5pt] (y) -- node[pos=0.6,anchor=north east]{$\,\termcolor\scriptstyle{i}$}  
	  (2.5,0.5) .. controls (a) and (b) .. (3,0.7) -- (3,2.6) -- (1,2.6) -- (x);
	\end{scope}
\end{tikzpicture}}
\]
\end{example}

The type system we introduce in this paper, simplified for this introduction to omit inputs and return values, records the probability of terminating with each given choice. A term $\term M$ is typed as follows, indicating that it may choose $i_k$ with probability $p_k$ from the choices $i_1$ through $i_n$. For simplicity, we restrict ourselves to rational probabilities, $p,q\in[0,1]\cap\mathbb{Q}$.
\[
	\term{M : p_1 i_1 * \cdots * p_n i_n }
\]
The type is a probability (sub-)distribution over the choices $\{i_1,\dots,i_n\}$, written as a formal sum. If the sum probability over $p_1$ through $p_n$ is one, the term is AST. For greater generality, and since our formalism readily incorporates it, we allow probabilities to sum to less than one, with the remainder representing an upper bound to the probability of divergence. We note here that, as a $\lambda$-calculus, the full calculus will have two means of non-termination: iteration with probability one, and higher-order non-terminating terms such as $\Omega=(\lambda x.xx)(\lambda x.xx)$. The type system will address both, and a term is almost-surely terminating if the sum probability of its type is exactly one.

\begin{example}
The body of our example term, without the loop, has the following type.
\[
	\term{\heads + (\tails + \,i) : /12\heads * /14\tails * /14i}
\]
\end{example}

Iteration $\term{M^i}$ removes the choice $i$ from the possible return choices. In the probabilistic setting, the remaining probabilities must then be renormalized to form a distribution. That is, if $i$ has probability $q$ (with $q<1$), the remaining choices (including divergence) have a total probability of $1-q$, and must be multiplied by $\frac1{1-q}$ to return the total to one. Typing for iteration is then as follows.
\[
\begin{array}{@{}r@{}l@{}}
	\term{M}   &{}: \type{p_1i_1 * \dots * p_ni_n * qj}
\\ \hline \\[-10pt]
	\term{M^j} &{}: \type{/{p_1}{1-q}i_1 * \dots * /{p_n}{1-q}i_n}
\end{array}	
\qquad
\term{M}=\looppic1 \qquad \term{M^j}=\looppic0
\]

\begin{example}
The type for our example term is given by taking the type $\type{/12\heads * /14\tails * /14i}$ of the body of the loop, removing the element $\type{/14i}$, and then multiplying the remaining sub-distribution $\type{/12\heads * /14\tails}$ by $\frac 43$ to renormalize it, which yields the following.
\[
	\term{(\heads + (\tails + \,i))^i : /23\heads * /13\tails}
\]
\end{example}

Types for iteration then correctly give the probabilities for each choice. Consider $\term{M^j}$ where $j$ has probability $q$ in $\term M$. For any other choice $i$ with probability $p$ in $\term M$ the chance of $\term{M^j}$ choosing $i$ is given by the following geometric series, generated by choosing $i$ immediately, after one iteration, after two iterations, and so forth. The limit of this series is the probability given by the type system.
\[
	p + pq + pq^2 + pq^3 + \cdots ~=~ \frac p{1-q}
\]


\subsubsection*{A Typed Term Calculus for Finite Markov Chains}

Further introducing the calculus, we first consider the fragment of \emph{ground terms}, as follows:
\[
	\text{\emph{Ground terms:}}\qquad \term{M,N}~\coloneqq~\term i~\mid~\term{M;i->N}~\mid~\term{M^i}~\mid~\term{?a.a}
\]
The first three are the \emph{control} constructs, introduced in \cite{Heijltjes-2025-MFPS}, which implement branching and iterating computation. The first (a \emph{choice} $\term i$) and the third (a \emph{loop} $\term{M^i}$), we have seen. The second is a \emph{case} $\term{M;i->N}$ which first evaluates $\term M$, and if $\term M$ terminates with $i$ then continues with $\term N$, otherwise discarding it. The control constructs reduce by the following rules:
\[
	\term{i;i->M}~\rw~\term{M} \qquad\quad \term{j;i->M}~\rw~\term j\quad (i\neq j) \qquad\quad \term{M^i}~\rw~\term{M;i->M^i}
\]
These constructions generalise sequential computation, which uses a single choice $\star$, with imperative $\emph{skip}$ as $\term*$ and \emph{sequencing} as $\term{M;N}=\term{M;*->N}$. The choice and case constructs are essentially \emph{throw} and \emph{try/catch} of exception handling, with exceptions $e$ as choices:
\[
	\mathsf{throw}~e~=~\term e \qquad\qquad \mathsf{try}~M~\mathsf{catch}~e~N~=~\term{M;e->N}
\]
Since cases are fall-through, they may be iterated to give a standard case switch. A conditional then encodes as below, where cases associate left, $\term{M; i->N ; j->P}=\term{(M; i->N) ; j-> P}$. We do need to keep constants $(\top,\bot)$ and exceptions $(e)$ apart: for the conditional below, if $\term M$ were to use $\bot$ as an exception, it would be caught by $\term N$. A while-loop, below that, is as follows: if $\term B$ returns $\bot$, then $\term M$ is skipped, the loop exited, and $\bot$ caught to return $\star$ (success); if $\term B$ returns $\top$, then $\term M$ is evaluated, iterating upon $\star$.
\[
\begin{array}{r@{}l}
	\mathsf{if}~B~\mathsf{then}~M~\mathsf{else}~N & ~=~\term{B;\top->M;\bot->N}
\\[5pt]	\mathsf{while}~B~\mathsf{do}~M            & ~=~\term{(B;\top->M)^\star;\bot->*}
\end{array}
\]
The fourth construct $\term{?a.a}$ is the \emph{generator} $\term{?a.M}$ of the probabilistic event $\lambda$-calculus~\cite{DalLago-Guerrieri-Heijltjes-2020} restricted to a variable $\term a$. The generator reduces as below: it flips a coin and binds the result, $\top$ or $\bot$, to the variable $\term a$ in $\term M$.
\[
	\term{?a.M}~\rw~\term{\{\top/a\}M}~\oplus~\term{\{\bot/a\}M}
\]
The restricted construct $\term{?a.a}$ thus reduces to $\top$ or $\bot$ with equal probability. Using it, a probabilistic sum encodes as follows.
\[
	M\oplus N~=~\term{?a.a;\top->M;\bot->N}
\]

\begin{example}
Our running example term unfolds as a ground term as follows.
\[
	\term{(\heads+(\tails+\,i))^i}~=~\term{`{\big(}\,?a.a ; \top->\heads ; \bot->(\,?b.b ; \top->\tails ; \bot -> i)`{\big)}^i}
\]
\end{example}

\begin{remark}
The generator $\term{?a.M}$ represents a \emph{fair} coin toss, with equal probability for $\top$ and $\bot$. We write $\term{?a_p.M}$ for an unfair coin with probability $p$ for $\top$ and $1-p$ for $\bot$. This may either be considered a primitive construct generalising $\term{?a.M}$, or constructed from fair coin tosses with iteration in the way our running example term generates $\frac23$ and $\frac13$.
\end{remark}

\begin{figure}
\[
\begin{array}{c@{\qquad}c@{\qquad}c@{\qquad}c}
  \infer{\term{i:1i}}{}
& \infer{\term{M;i->N: p_I * (p_J+q.`r\!_J) * q.`r\!_K}}{\term{M: p_I * p_J * qi} && \term{N: `r\!_J * `r\!_K}}
& \infer{\term{M^i:/1{1-q}.p_I}}{\term{M:p_I * qi} & (q\neq 1)}
& \infer{\term{?a.a: /12\top * /12\bot}}{}
\\ \\
\vc{\begin{tikzpicture}[thick,x=18pt,y=15pt,outer sep=0pt,inner sep=1pt,->,>=stealth]

	\coordinate (m) at (0,0);
	
	\node (i) at (0,-2) {$\type{1i}$};
	
	\draw (m)--(i.north);
	
	\node[draw=red,fill=red!10,rounded corners,minimum size=16pt] (M) at (m) {$\term i$};
	\draw (0,2)--(M);
\end{tikzpicture}}
&
\vc{\begin{tikzpicture}[thick,x=24pt,y=12pt,outer sep=0pt,inner sep=1pt,->,>=stealth]
	\coordinate (m) at (0,0);
	\coordinate (n) at (1,-2);
	\node[minimum size=16pt] (n') at (n) {};
	
	\node (I) at (-1  ,-4) {$\type{p_I}\vphantom)$};
	\node (J) at ( 0  ,-5.5) {$\type{p_J+q.`r\!_J}\vphantom)$};
	\node (K) at ( 1  ,-4) {$\type{q.`r\!_K}\vphantom)$};
	
	\draw[double] (m)--(-1,-2)--(I.north);
	\draw[double] (m)--( 0,-4) (n)--( 0,-4)--(J.north);
	\draw[double] (n)--(K.north);

	\draw (0,2)--(m);
	\draw (m)--node[pos=0.6,anchor=south west]{$\type{qi}$} (n'.north);
	
	\node[draw=red,fill=red!10,rounded corners,minimum size=16pt] (M) at (m) {$\term M$};
	\node[draw=red,fill=red!10,rounded corners,minimum size=16pt] (N) at (n) {$\term N$};
\end{tikzpicture}}
&
\vc{\begin{tikzpicture}[thick,x=18pt,y=15pt,outer sep=0pt,inner sep=1pt]
	\coordinate (m) at (0,0);
	
	\node[outer sep=2pt] (I) at (-0.5,-2) {$\type{/1{1-q}.p_I}$};
	
	\draw[double,->,>=stealth] (m)--(I.north) ;
	
	\coordinate (a) at (0.85,-1.7);
	\coordinate (b) at (1.25,-1.6);
	
    \draw[rounded corners=5pt] (m) 
       -- (0.5,-1) .. controls (a) and (b) .. (1.25,-.75) 
       --node[pos=0.3,anchor=west]{\,$\type{qi}$} (1.25,1.5) -- (0,1.5) -- (0,1) -- (m);

	\node[draw=red,fill=red!10,rounded corners,minimum size=16pt] (M) at (m) {$\term M$};
	\draw[->,>=stealth] (0,2)--(M);	
\end{tikzpicture}}
&
\vc{\begin{tikzpicture}[thick,x=18pt,y=15pt,outer sep=0pt,inner sep=1pt,->,>=stealth]

	\coordinate (m) at (0,0);
	
	\node[outer sep=2pt] (i) at (-0.8,-2) {$\type{/12\top}$};
	\node[outer sep=2pt] (j) at ( 0.8,-2) {$\type{/12\bot}$};
	
	\draw (m)--(i.north) ;
	\draw (m)--(j.north) ;
	
	\node[draw=red,fill=red!10,rounded corners,minimum height=18pt,minimum width=30pt] (M) at (m) {};
	\node at (m) {$\term{?a.a}$};
	\draw (0,2)--(M);	
\end{tikzpicture}}
\end{array}
\]
\caption{Typing rules for ground terms, with illustrating diagrams. The double arrows represent multiple branches. In the second rule, the types for $\term M$ and $\term N$ divide into overlapping parts $\type{p_J}$ and $\type{`r\!_J}$, and non-overlapping parts $\type{p_I}$ and $\type{`r\!_K}$, where $I$, $J$, and $K$ are pairwise disjoint sets of choices.}
\label{fig:ground types}
\end{figure}
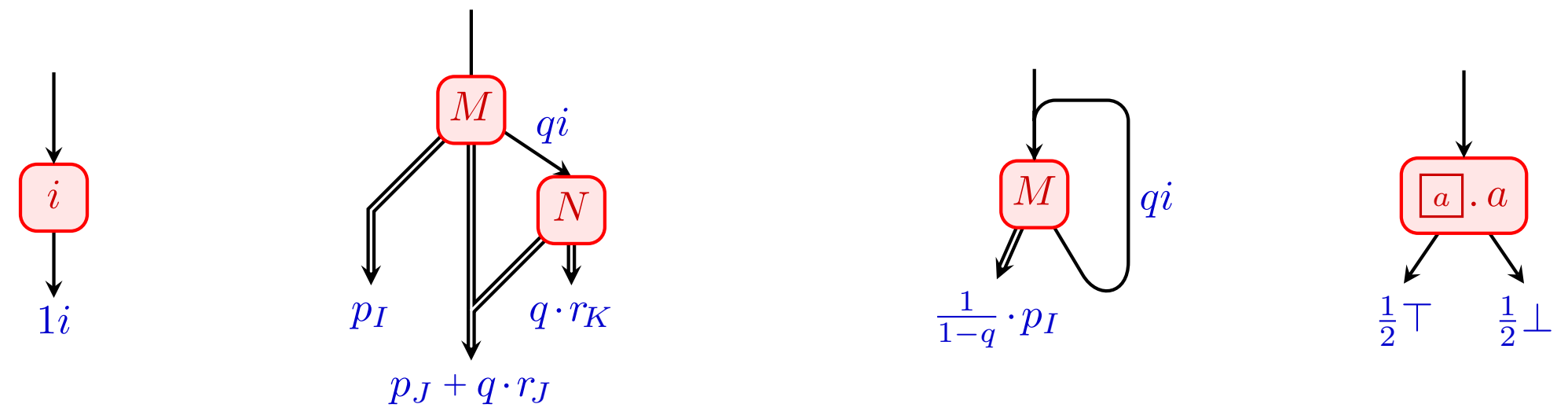

We now turn to types. Deterministic types for the control constructs record the possible terminating choices for a term, as is standard for exception handling. Types are finite sets of choices, written as formal sums $\type I=\type{i_1+\cdots+i_n}$; semantically, these are coproducts. We have the following typing rules:
\[
            \infer{\term{i:i+I}}{} 
\qquad\quad	\infer{\term{M;i->N : I\cup J}}{\term{M:I+i} && \term{N:J}}
\qquad\quad \infer{\term{M^i : I}}{\term{M:I+i}}
\]
As in the previous, in the probabilistic setting types become (sub-)distributions $\type{p_I}=\type{p_1i_1 * \cdots * p_ni_n}$ over a set $I$ of choices. Ground terms are then typed by the rules in Figure~\ref{fig:ground types}, where we write $\type{p_I+q_I}$ for the pointwise addition of two distributions over $I$ and $\type{p.q_I}$ for the multiplication of each probability in $\type{q_I}$ by $\type p$. A type $\type{p_I}$ implicitly assigns probability zero to any choice not in $I$, corresponding to the weakened $\type I$ in the first rule above. Observe that the probabilistic type system is a direct generalization of the deterministic one.

In Section~\ref{sec:ground terms} we will return to ground terms and prove a \emph{Kleene theorem} for them: ground terms are equivalent to finite Markov chains, given as probabilistic automata, where states correspond to choices. Then types compute the return distribution of an encoded Markov chain, while type inference is decidable, and terms typed with total distributions are positive AST.


\subsubsection*{The Probabilistic FMC}

The FMC combines the control constructs with the $\lambda$-calculus via the Krivine abstract machine (KAM)~\cite{Krivine-2007}, which evaluates terms in the context of an argument stack: application $M\,N$ pushes $N$ and continues as $M$; abstraction $\lambda x.M$ pops (say) $N$ and continues as $\{N/x\}M$. This gives a canonical interpretation to the (call--by--name) $\lambda$-calculus in terms of \emph{stateful} and \emph{sequential} computation, where the stack is the state and computation proceeds sequentially via \emph{pop-} and \emph{push-}actions. Together with the control constructs, the FMC then features branching and iterating computation that inputs and outputs a stack.

Emphasizing this perspective, the FMC uses $\term{[N].M}$ for application $M\,N$ and $\term{<x>.M}$ for abstraction $\lambda x.M$. With the control constructs and generator as above, the \emph{probabilistic FMC} is as follows.
\[
\term{M,N} ~\coloneqq~ \term{x} ~\mid~ \term{<x>.M} ~\mid~ \term{[N].M} ~\mid~ \term{i} ~\mid~ \term{M;i->N} ~\mid~ \term{M^i} ~\mid~ \term{?a.M}
\]

\begin{example}
\label{ex:ho-term}
Consider the following example term.
\[
	\term{<x>.x~;~\heads -> <y>.* ~;~ \tails -> <y>.[y].[y].*}
\]
It takes a (fair or unfair) coin from the stack as $\term x$, such as our $\frac13$ probability running example term, and tosses it. On heads, it deletes the top element on the stack; on tails, it duplicates it. Both branches return with the default choice $\term *$ (``skip'').
\end{example}

\begin{remark}
\label{rem:CBPV}
The notion of sequencing in the FMC is fundamentally the same as that of Moggi's computational metalanguage~\cite{Moggi-1991} and Levy's call--by--push--value~\cite{Levy-2003}: 
\[
	M\bind N ~=~ \term{M;N}
\]
The difference is that the latter return a single value, while the FMC returns the whole stack: $\mathsf{return}~M~=~\term{[M].*}$. This notion of sequencing further allows the encoding of call--by--value behaviour, both in a deterministic and probabilistic setting; see~\cite{Heijltjes-2023}, respectively~\cite{Heijltjes-Majury-2025}.
\end{remark}

For the full calculus, the distribution types of ground terms are extended with types for the input stack and for the return stack of each branch. Types (or \emph{computation types}) $\type{t}$ and stack types $\type{!t}$ are given below; the latter are vectors giving a type for each term, and the former an implication from an input stack type $\type{!s}$ to a formal sum of triples $\type{!t.pi}$ of an output stack type $\type{!t}$, a probability $p$, and a choice $i$, where the restriction to $\type{pi}$ forms a (sub-)distribution.
\[
\begin{array}{l@{\qquad}r@{}l}
        \text{\emph{Types:}}	    & \type{s,t}~\coloneqq~{} & \type{!s ~=>~ !t_1.p_1i_1 * \cdots * !t_n.p_ni_n}
\\[5pt]	\text{\emph{Stack types:}}  & \type{!t}~\coloneqq~{}  & \type{t_1..t_n}
\end{array}
\qquad
\vc
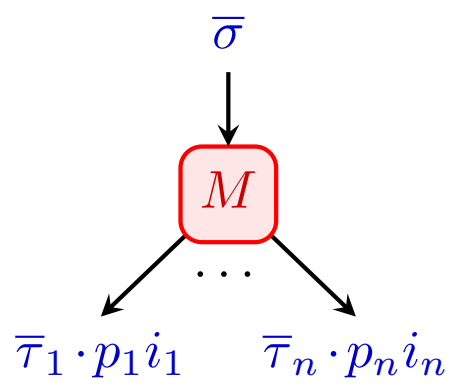
{\begin{tikzpicture}[thick,x=24pt,y=15pt,outer sep=0pt,inner sep=1pt,->,>=stealth]

	\coordinate (m) at (0,0);
	
	\node[outer sep=1pt] (r) at ( 0, 2) {$\type{!s}$};
	\node[outer sep=1pt] (i) at (-1,-2) {$\type{!t_1.p_1i_1}$};
	\node[outer sep=1pt] (j) at ( 1,-2) {$\type{!t_n.p_ni_n}$};
	\node at (0,-1) {$\dots$};
	
	\draw (m)--(i.north) ;
	\draw (m)--(j.north) ;
	
	\node[draw=red,fill=red!10,rounded corners,minimum size=18pt] (M) at (m) {$\term{M}$};
	\draw (r)--(M);
\end{tikzpicture}}
\]
Since stacks are last-in-first-out, identity terms are of the form $\term{<x>.<y>.[y].[x].*}$, reversing the order of variables (and returning the default choice $\star$). Type vectors follow the order of terms, implicitly reversing on the left of an implication, so that identity terms have type $\type{!t => !t.1\star}$; for the given example, $\term{<x>.<y>.[y].[x].* : s\,t => t\,s.1\star}$. Types for ground terms as distributions abbreviate computation types with empty type vectors, written $\type{\e}$.
\[
	\type{p_1i_1*\cdots*p_ni_n}\quad=\quad \type{\e~=>~\e.p_1i_1*\cdots*\e.p_ni_n}
\]

\begin{example}
In the sequential setting, with a single branch $\star$, composition of two terms $\term{M:!r=>!s.p\star}$ and $\term{N:!s=>!t.q\star}$ multiplies their probabilities, $\term{M;N : !r => !t.(pq)\star}$.
\end{example}

\begin{example}
The term of Example~\ref{ex:ho-term} cannot be typed because it features two branches with the same choice $\term*$ but different output stacks (one empty, the other holding two copies of the input $\term y$). Adapting it to use distinct choices $i$ and $j$ instead, and choosing the type for the first input to match our running example, we get the type below (for any type $\type t$ for $\term y$). 
\[
	\term{<x>.x~;~\heads -> <y>.i ~;~ \tails -> <y>.[y].[y].j ~:~ (\e=>\e./23\heads*\e./13\tails)\,t~=>~\e./23i~*~\,t\,t./13j }
\]
The term expects two inputs: the unfair coin and some argument $\type t$ for the abstractions $\term{<y>}$ of each branch (note that since both branches pop from the same input stack, both variables $\term y$ get the same type). It then produces two possible outputs: with probability $\frac 23$ the branch $i$ with the empty stack, and with probability $\frac 13$ the branch $j$ with two copies of $\term y$ on the stack.
\end{example}

As the example demonstrates, the type system is quite restrictive: probabilities are fixed, even on inputs. However, the potential of lifting these restrictions is clear, for example by introducing polymorphism (variables) for probabilities, or generalising to dependent types. The purpose of the present paper is not just to establish the basic system of probabilistic simple types for AST, but to offer a clear way forward in probabilistic type systems.

\begin{remark}
\label{rem:monads}
Our distribution types $\type{!t_1.p_1i_1 * \cdots * !t_n.p_ni_n}$ are a fundamentally different construct than a \emph{probability monad} $P(A)$~\cite{Giry-1982}: the former features probability distributions at the \emph{type level}, the latter at the \emph{term level}. Since sampling is a basic construct, terms in the probabilistic FMC are inherently probabilistic in the way of a probability monad; for example, assuming a base type $\type\Z$ of integers, the following term returns $2$ or $3$ with equal probability, despite having only a trivial probability at the type level.
\[
	\term{?a.a~;~\top -> [2].*~;~\bot -> [3].*~:~\e => \Z.1\star} 
\]
\end{remark}

\begin{remark}
The Functional Machine Calculus as presented in~\cite{Heijltjes-2023,Heijltjes-2025-MFPS} allows not just one, but multiple argument stacks, which model various state-like effects: mutable store (as a stack of depth one), input/output (as dedicated input-only and output-only stacks), and probabilistic choice (as a probabilistically-generated input stack); see op.cit.\ for a detailed exposition. The probabilistic FMC is a proper fragment with two stacks: one for the $\lambda$-calculus, and a probabilistic one, viewing the generator construct $\term{?a.M}$ as popping a value from this stack as $a$. The same technique, sampling from a random stream on the Krivine machine, is used by Ehrhard~\cite{Ehrhard-2019}.
\end{remark}

After exploring related work in the next section, we will define the calculus (Section~\ref{sec:PFMC}) and the type system (Section~\ref{sec:types}), and give our main result, Theorem~\ref{thm:termination}: the probability given by the type of a term is a lower bound on the probability of termination, which implies almost-sure termination if that probability is one (Section~\ref{sec:AST}). The final sections explore the expressiveness of the calculus, demonstrating that ground terms are equivalent to finite Markov chains and thus positive AST (Section~\ref{sec:ground terms}), and that the full calculus is \emph{null AST}: while the probability of termination is one, expected runtime is infinite (Section~\ref{sec:NAST}).


\section{Related Work}
\label{sec:related work}

Our type system is a conservative extension of our previous systems in \cite{Antonelli-DalLago-Pistone-2022} and \cite{Heijltjes-Majury-2025} (see Remark~\ref{rem:probabilistic types embed}). While those systems already feature higher-order functions, they do not account for either iteration or recursion.

The literature on semantics and termination properties of probabilistic computation is extensive and includes a wide range of techniques. We mention semantic domains~\cite{Dahlqvist-Kozen-2020,Kozen-1981,Vakar-Kammar-Staton-2019}, bisimulations and trace semantics~\cite{vanGlabbeek-Smolka-Steffen-1995,Silva-Sokolova-2011,Stark-Smolka-2000}, probabilistic automata~\cite{Esparza-Kucera-Mayr-2004,Etessami-Yannakakis-2009}, probabilistic coherence spaces~\cite{Crubille-Ehrhard-Pagani-Tasson-2017,Danos-Ehrhard-2011,Ehrhard-Tasson-2019}, program  logics~\cite{FerrerFioriti-Hermanns-2015,McIver-Morgan-Kaminski-Katoen-2018}, and separation logic~\cite{Barthe-Hsu-Liao-2020,Jereb-Simpson-2025}.

Type systems for probabilistic calculi are numerous and diverse as well. We distinguish two kinds. Firstly, there are deterministic or monadic types, which do not feature explicit probabilities at the type level (see Remark~\ref{rem:monads}). These tend to play a supporting role, with probabilistic properties such as AST then guaranteed by construction. Examples are~\cite{TorrezRuiz-Piedeleu-Silva-Zanasi-2024},~\cite{Staton-2017} which features coproduct-typed (but not probabilistic) iteration, and~\cite{Breuvart-DalLago-Herrou-2021} which extends G\"odel's $\mathbb{T}$ with probabilistic generators and iterators that are inherently AST.

Secondly, there are several formal methodologies enforcing AST or PAST in probabilistic $\lambda$-calculi with recursion. These are generally highly expressive, but complex, using involved techniques such as refinement types~\cite{Avanzini-DalLago-Ghyselen-2019,Kura-Unno-2024}, size types~\cite{DalLago-Grellois-2019}, non well founded type systems~\cite{DalLago-Padovani-2024,Das-Wang-Hoffmann-2023}, intersection types~\cite{Beutner-Ong-2021,DalLago-Faggian-Ronchi-2021}, guarded refinements~\cite{Gregersen-Aguirre-Haselwarter-Tassarotti-Birkedal-2024}, or CPS transformations~\cite{Avanzini-Barthe-DalLago-2021}. Such complexity is largely unavoidable: unlike in the deterministic case, probabilistic termination is undecidable even for higher-order programs operating over finite datatypes~\cite{Kobayashi-DalLago-Grellois-2020}, as exemplified by probabilistic variants of the $\lambda Y$-calculus and higher-order recursion schemes. Expressiveness thus comes at the expense of simplicity; here, we prioritize simplicity, obtaining a system that constitutes a minimal extension of simple types.

Our approach sets itself apart in two ways. Firstly, from the FMC we inherit the combination of higher-order functions with \emph{iteration}, rather than \emph{recursion} (we comment on typing recursion in our system in Appendix~\ref{A:recursion}). Secondly, as a direct probabilistic generalisation of coproducts, our probabilistic types are a notion of \emph{simple types}. We are thus taking a ``bottom-up'' approach, consciously trading expressiveness for simplicity and extensibility: our simple types are highly restricted in the algorithms they may express, but they arise naturally from typed iteration, and we expect they will extend to more powerful type systems such as polymorphic types, intersection types, and dependent types along standard lines.

It would be interesting to consider continuation-passing translations from our calculus into one with recursion. Generally, these require higher types; for instance, the simply-typed FMC, when translated this way into $\lambda$-calculus, requires System F types. With such a translation for the probabilistic FMC, it would then become possible to compare the expressive power of our system against the approaches  mentioned above. This is however out of scope for the present publication.


\section{The Probabilistic FMC}
\label{sec:PFMC}

For the formal development, we first recall the terms of the probabilistic FMC from the introduction, and then give its operational semantics. The FMC further features a confluent reduction relation~\cite{Heijltjes-2025-MFPS}, which we omit here.

\begin{definition}
The \emph{terms} of the \PFMC\ are given by the following grammar.
\[
\term{M,N} ~\coloneqq~ \term{x} ~\mid~ \term{<x>.M} ~\mid~ \term{[N].M} ~\mid~ \term{i} ~\mid~ \term{M;i->N} ~\mid~ \term{M^i} ~\mid~ \term{?a.M}
\]
The constructs are: a \emph{variable} $\term x$; a \emph{pop} $\term{<x>.M}$ that binds $\term x$ in $\term M$; a \emph{push} $\term{[N].M}$; a \emph{choice} $\term i$; a \emph{case} $\term{M;i->N}$; a \emph{loop} $\term{M^i}$; and a \emph{generator} $\term{?a.M}$ which binds the variable $\term a$ in $\term M$.
\end{definition}


The small-step semantics is given by the Krivine machine~\cite{Krivine-2007}, which evaluates the $\lambda$-calculus via a single argument stack, with the following extensions. The generator $\term{?a.M}$ will pop from a stack of boolean values, called a \emph{trace}, which we will model as being probabilistically generated. Branching computation is implemented via a \emph{continuation stack} of conditional continuations $\term{i->N}$, where $\term{M;i->N}$ pushes the continuation $\term{i->N}$, and a choice term $\term i$ discards non-matching continuations $\term{j->M}$ where $i\neq j$ until it pops a matching $\term{i->N}$. The machine terminates successfully when it reaches a choice $\term i$ and the continuation stack is empty, at which point it returns the argument stack.

\begin{definition}
The \emph{functional abstract machine} is given by the following data. \emph{Traces} $r,s,t$, \emph{argument stacks} $R,S,T$, and \emph{continuation stacks} $K,L$ are as follows.
\[
    r,s,t ~\coloneqq~ \e ~\mid~ \bot\,t ~\mid~ \top\,t   \qquad\quad 
    R,S,T ~\coloneqq~ \e ~\mid~ S\,\term M               \qquad\quad 
    K,L   ~\coloneqq~ \e ~\mid~ (\term{i->M})\,K
\]
The \emph{states} of the machine are four-tuples $(t,S,\term M,K)$. Its \emph{steps} or \emph{transitions} are given in Figure~\ref{fig:machine}. A \emph{run} is a sequence of steps, written as below. A run is \emph{successful} if it terminates in a \emph{final state} $\state\e Ti\e$ with a choice term $\term i$ and empty trace and continuation stack.
\[ 
	\step= sSMK tTNL
\]
\end{definition}

\noindent
Appendix~\ref{A:run} demonstrates a successful run of the machine for our running example term.

\begin{figure}
\[
\begin{array}{lccr}
	  \mathsf{push}     & \step- t S {[N].M} K             t {S\,\term N} M K
	                    & \step- t S i {(\term{i->M})\,K}  t S M K                    & \mathsf{select}
\\ \\ \mathsf{pop}      & \step- t {S\,\term N} {<x>.M} K  t S {\{N/x\}M} K
	                    & \step- t S i {(\term{j->M})\,K}  t S i K                    & (i\neq j)\quad \mathsf{reject}
\\ \\ \mathsf{sequence} & \step- t S {M;i->N} K            t S M {(\term{i->N})\,K}
	                    & \step- t S {M^i} K               t S M {(\term{i->M^i})\,K} & \mathsf{iterate}
\\ \\ \mathsf{sample}   & \step- {\top\,t} S {?a.M} K      t S {\{\top/a\}M} K
                        & \step- {\bot\,t} S {?a.M} K      t S {\{\bot/a\}M} K
\end{array}
\]
\caption{The Abstract Machine}
\label{fig:machine}
\end{figure}

In addition to the machine we give a big-step operational semantics, in the form of an inductively defined evaluation relation $(\evalarrow)$ that describes the successful runs of the machine directly. Where the machine implements branching computation via the contination stack, the big-step semantics gives a more abstract view as conditional sequential composition.

\begin{definition}
The \emph{evaluation} relation has statements $\eval tSM Ti$ and is defined inductively by the following rules.
\[
\begin{array}{ccc}
     \infer[\Rax]  {\eval \e Si Si}{}
   & \infer[\Rsel] {\eval {rs}R{M;i->N} Tj }{\eval rRM Si & \eval sSN Tj}
   & \infer[\Rrej] {\eval rR{M;i->N} Sj }{\eval rRM Sj & (i\neq j)}
\\\\ \infer[\Rpush]{\eval sS{[N].M} Ti}{\eval s{S\,\term N}M Ti}
   & \infer[\Rloop]{\eval {rs}R{M^i} Tj }{\eval rRM Si & \eval sS{M^i} Tj}
   & \infer[\Rexit]{\eval rR{M^i} Sj }{\eval rRM Sj & (i \neq j)} 
\\\\ \infer[\Rpop] {\eval t{S\,\term N}{<x>.M} Ti}{\eval tS{\{N/x\}M} Ti}
   & \infer[\Rbot] {\eval {\bot\,t}S{?a.M} Ti}{\eval tS{\{\bot/a\}M} Ti}
   & \infer[\Rtop] {\eval {\top\,t}S{?a.M} Ti}{\eval tS{\{\top/a\}M} Ti}
\end{array}
\]
\end{definition}

\noindent
The machine and the big-step evaluation relation can be put in correspondence as follows.

\begin{proposition}[{\cite[Proposition 5.4]{Heijltjes-2025-MFPS}}]
Evaluation characterizes successful machine termination:
\[
\eval tSMTi \quad \iff \quad \step= tSM\e \e Ti\e
\]
\end{proposition}


The calculus as defined above is deterministic: it externalises the probabilistic component of evaluation to the input trace $t$. To give a probabilistic semantics, we will define an evaluation relation $\Eval SM\DT$ which takes a stack $S$ and a term $\term M$ to a certain kind of distribution $\DT$, a \emph{return family}. This is a set of traces with for each $t$ an associated pair 
$(T,i)$ of a return stack and choice such that $\eval tSMTi$ (effectively, we are only shifting where the trace $t$ is considered as a parameter, on the machine versus in the return distribution). Since the probability of a trace $t$ is $\frac12^n$ where $n$ is the length of $t$, a return family readily gives rise to a probability (sub-)distribution.

A return family collects evaluation results for distinct probabilistic choices, not further evaluation along the same trace: if it has $t$, it should not have any prefix of $t$, nor any trace extending $t$. A return family will therefore be defined over a \emph{frontier} $F$: a (possibly infinite) set of traces such that no element in $F$ is a prefix of another. The \emph{probability} $\pr F$ of a frontier $F$ is the sum probability of its traces; observe that by definition, $\pr F\leq 1$.
\[
\begin{array}{l@{}l}
	    \pr t &~=~ (\frac12)^{|t|} \quad\text{where $|t|$ is the length of $t$}
\\[5pt]	\pr F &~=~ \sum_{t\in F}\pr t
\end{array}
\]

\begin{example}
\label{ex:frontier}
The set of traces $\{ (\bot^n)\top\mid n\in \N \} = \{ \top,\bot\top,\bot\bot\top,\dots\}$ is a frontier with probability $1$.
\end{example}

A \emph{return family} $\DS$ then collects a return stack and choice, as a pair $(S_t,i_t)_t$, for every trace $t$ in a frontier $F$, as below. The frontier $F$ over which $\DS$ is defined is written $\floor\DS=F$, and the \emph{probability} of a return family is that of its frontier, $\pr\DS = \pr{\floor\DS}$.
\[
	\DS,\DT~\coloneqq~\{ (S_t,i_t)_t \}_{t\in F}
\]

\begin{definition}
A \emph{probabilistic evaluation} $\Eval SM\DT$ of a term $\term M$ and stack $S$ to a return family $\DT$ is given by the evaluation of $\term M$ and $\term S$ with each $t\in F$, as follows.
\[
	\Eval SM{\{ (T_t,i_t)_t \}_{t\in F}} \iff \forall t\in F.~(\eval tSM {T_t}{i_t})
\]
\end{definition}

\begin{example}
For our running example term we have the evaluation $\Eval \e {(\heads + (\tails + \,i))^i} \DT$ for the following return family $\DT$ over the frontier of Example~\ref{ex:frontier}.
\[
	\DT~=~\{ (\e,\heads)_t \mid t = (\bot^{2n})\top,~n\in\N~\} \cup \{ (\e,\tails)_t \mid t = (\bot^{2n+1})\top,~n\in\N~\}
\] 
\end{example}

Observe that evaluation is closed under inclusion: if $\Eval RM\DT$ and $\DS\subseteq\DT$ then $\Eval RM\DS$. An evaluation $\Eval SM\DT$ gives a lower bound of $\pr{\DT}$ for the probability of termination of a term $\term M$ in the context of a stack $S$.

\begin{definition}
The \emph{probability of termination} of a term $\term M$ in the context of a stack $S$ is the largest probability $\pr{\DT}$ for any $\DT$ such that $\Eval SM\DT$.
\end{definition}

To work with return families we use the following notation. The \emph{sum} of two return families $\DS+\DT$ is defined if they have disjoint frontiers, $\floor\DS\cap\floor\DT=\varnothing$, and is the union of their underlying sets $\DS\cup\DT$. Likewise, \emph{inclusion} between families $\DS\subseteq\DT$ is inclusion as sets. We define the following notions below. On the left, $s\pfx F$ and $s\pfx\DS$ add $s$ as a prefix to every trace $t$ in $F$, respectively in $\DS$, and $S\,\DT$ prefixes every stack in $\DT$ with $S$. On the right, $\DS\only i$ restricts a family to the choice $i$ while $\DS\less i$ excludes the choice $i$, and $\stk\DS$ collects the stacks in $\DS$ as a set. 
\[
\begin{array}{r@{}l@{\qquad\qquad}r@{}l}
		s\pfx F    &{}~=~\{ \, st \mid t\in F \,\}                & \DS\only i &{}~=~\{ \,(S,j)_t \in \DS \mid j=i \,\} 
\\[5pt] s\pfx\DS   &{}~=~\{ \,(S,i)_{st} \mid (S,i)_t \in \DS\,\} & \DS\less i &{}~=~\{ \,(S,j)_t \in \DS \mid j\neq i \,\}
\\[5pt] S\,\DT     &{}~=~\{ \,(S\,T,i)_t \mid (T,i)_t \in \DT\,\} & \stk\DS    &{}~=~\{ \,S \mid (S,i)_t \in \DS\,\}
\end{array}
\]
We will freely use the following equations.
\[
	\floor{s\pfx\DS} = s\pfx\floor\DS    \qquad\quad
	\pr{s\pfx\DS}    = \pr s\times\pr\DS \qquad\quad
	\pr{\DS+\DT}     = \pr\DS + \pr\DT
\]


\section{Types}
\label{sec:types}

Types for the FMC describe the input/output behaviour of the machine. In the deterministic case, a type $\type t$ is of the form below left, where the type vector $\type{!s}=\type{s_1 .. s_k}$ types the input stack on the machine, and the type vectors $\type{!t_m}$ type the return stack for each choice $i_m$. The type indicates that given an input stack $S:\type{!s}$ the machine will return some choice $i_m$ and stack $T:\type{!t_m}$. For the probabilistic case we replace the formal sum of the return type with a probability distribution, adding a probability $p_m$ for each summand as below right.
\[
	\type{!s ~=>~!t_1.i_1 + .. + !t_n.i_n} \qquad\qquad\qquad
	\type{!s ~=>~!t_1.p_1i_1 * .. * !t_n.p_ni_n}
\]
We will formalize the return type as a \emph{distribution type}, a family in a finite set $I$ of choice labels, with a type vector $\type{!t_i}$ and a probability $p_i$ for each $i\in I$ such that the sum probability over all $p_i$ is at most one. It is thus a combination of a deterministic sum type (the family of type vectors $\type{!t_i}$) and a probability (sub-)distribution over $I$ (the family of probabilities $p_i$).

\begin{definition}
\emph{Types}, \emph{stack types}, and \emph{distribution types} are given as follows.
\[
\begin{array}{l@{\qquad}r@{~\coloneqq~}l}
	\text{types}	            & \type{s,t} & \type{!t => d}
\\	\text{stack types}          & \type{!t}  & \type{t_1..t_n}
\\  \text{distribution types}   & \type{g,d} & \type{\{(!t_i,p_i)_i\}_{i\in I}}\quad\text{where}\quad \sum_{i\in I}p_i\leq 1
\end{array}
\]
\end{definition}

Recall from the introduction that stack types on the left of an implication $\type{=>}$ are implicitly reversed. Distribution types may implicitly expand with zero-probability summands when required by the typing rules (we will highlight the relevant cases in the discussion of the typing rules).

\begin{remark}
\label{rem:simple types embed}
Our types are a conservative extension of simple types for the $\lambda$-calculus, which embed as follows (see also~\cite{Heijltjes-2023,Heijltjes-2025-MFPS,Heijltjes-Majury-2025}). The base type $o$ becomes $\type{\e=>\e.1\star}$, the type of the default choice $\term*$. The type arrow $\rho\to\tau$ becomes the operation of adding one further input type $\type r$ to the inputs of $\tau$; i.e.\ letting $\tau=\type{!s => d}$ then $\type{r \to (!s => d)}=\type{r\,!s => d}$. Together, we may view the embedding also as below. 
\[
	\typ{t_1 \to \dots \to t_n \to o} \quad=\quad \type{t_1 \dots t_n => \e.1\star}
\]
\end{remark}

We introduce the following notation to work with distribution types. The \emph{basis} $\floor{\type d}$ of a distribution type $\type d$ is the set of choices $\floor{\type d}=I$ over which it is defined. To recover the informal notation of the introduction, we define the following: the \emph{singleton} distribution type $\type{!t.pi}=\type{\{(!t,p)_i\}}$, the \emph{empty} distribution type $\type{0}$ with $\floor{\type 0}=\varnothing$, and the (disjoint) \emph{sum} $\type{g*d}=\type{g\cup d}$ of two distribution types $\type g$ and $\type d$ with no choice in common, $\floor{\type g}\cap\floor{\type d}=\varnothing$. The \emph{probability} $\pr{\type d}$ of a distribution type is the sum over its probabilities $p_i$. Below left, the notation $\type{!s\,d}$ prefixes the type vector $\type{!s}$ to each $\type{!t}$ in $\type d$, and $\type{q.d}$ multiplies each probability $p_i$ in $\type d$ by the probability $q$. Below right, $\type{d|I}$ restricts $\type d$ to the choices in $I$, and $\type{d-I}$ removes them.
\[
\begin{array}{r@{}l@{\qquad\qquad}r@{}l}
	    \type{q.d}   &{}~=~ \type{\{    (!t,qp)_i \mid (!t,p)_i\in d\, \}}  & \type{d|I} &{}~=~ \type{\{ (!t.p)_i \in d \mid i\in I\,\}}
\\[5pt]	\type{!s\,d} &{}~=~ \type{\{ (!s\,!t,p)_i \mid (!t,p)_i\in d\, \}}  & \type{d-I} &{}~=~ \type{\{ (!t.p)_i \in d \mid i\notin I\,\}}
\end{array}
\]
Finally, the \emph{merger} $\type{g++d}$ of two distribution types is defined if for every $i\in\floor{\type g}\cap\floor{\type d}$ they have the same type vector $\type{!t_i}$, and is as follows.
\[
	\type{g++d} ~=~ \type{ \{ (!t,p+q)_i \mid i\in {\floor\gamma}\cup{\floor\delta}~`,~(!t,p)_i \in g~`,~(!t,q)_i\in d\}~*~g-{\floor\delta}~*~d-{\floor\gamma} }
\]

\begin{definition}
The \emph{typed PFMC} (\PFMCT\!) is given by the typing rules in Figure~\ref{fig:types}, where a context $\term{G}=\term{x_1:t_1,,x_n:t_n}$ is a finite function from variables to types written as a sequence, and typing judgment $\term{G |- M:t}$ assigns a term $\term M$ the type $\type t$ in the context $\term G$.
\end{definition}

\begin{figure}
\[
\begin{array}{l@{\qquad}c@{\qquad}c@{\qquad}r}
      \mathsf{variable}  & \vc{\infer{\term{G, x:t |- x:t}}{}}
                         & \vc{\infer{\term{G |- M : \e => 0}}{}}                                     & \mathsf{zero}
\\ \\ \mathsf{pop}       & \vc{\infer{\term{G |- <x>.M : r\,!s => d}}{\term{G , x:r |- M: !s => d}}}
                         & \vc{\infer{\term{G |- [N].M : !s => d}}
                                     {\term{G |- N:r} && \term{G |- M : r\,!s => d}}}                 & \mathsf{push}
\\ \\ \mathsf{choice}    & \vc{\infer{\term{G |- i: \e => \e.1i}}{}}
                         & \vc{\infer{\term{G |- M;i->N : !r => g ++ (p . d)}}
                                     {\term{G |- M: !r => g * !s.pi} && 
                                      \term{G |- N: !s => d}}}                                        & \mathsf{sequence}
\\ \\ \mathsf{expand}    & \vc{\infer{\term{G |- M: !s\,!r => !r\,d}}{\term{G |- M: !s => d}}}
                         & \vc{\infer{\term{G |- M^i: !s => /1{1-p} . d}}
                                     {\term{G |- M: !s => d * !s.pi} && (p\neq 1)} }                  & \mathsf{iterate}    
\\ \\ \mathsf{sample}    & \multicolumn{2}{c}{
                           \vc{\infer{\term{G |- ?a.M : !r => (/12 . g) ++ (/12 . d)}}
                                     {\term{G , a: \e => \e.1\top |- M: !r => g} && 
                                      \term{G , a: \e => \e.1\bot |- M: !r => d}}}
                          }
\end{array}
\]
\caption{The typed probabilistic FMC}
\label{fig:types}
\end{figure}

We will first discuss the typing rules in detail, skipping the \emph{variable} rule as it is standard.

\begin{itemize}
	\item 
The \emph{zero} rule allows to type any term $\term M$ with a type $\type{\e =>0}$, returning the empty sub-distribution, which has sum probability zero. This is necessary to type terms such as $\term{*+\Omega}$, which returns immediately or diverges with equal probability, where we type $\term\Omega$ with the zero rule.

	\item
The \emph{pop} and \emph{push} rules add, respectively remove, the first stack input. With the embedding of the type arrow $\type{\to}$ of the simply-typed $\lambda$-calculus by $\type{r \to (!s => d)}=\type{r\,!s => d}$, these are the standard rules for abstraction and application.

	\item
The \emph{choice} rule expresses that $\term i$ immediately returns for the empty stack. Using the \emph{expand} rule it may be typed $\type{!s => !s.1i}$, reflecting that it returns any input stack unchanged.

	\item
The \emph{sequence} rule reflects that $\term{M;i->N}$ first evaluates $\term{M}$, with input $\type{!r}$ and output either $\type{!s}$ on branch $i$ with probability $p$, or a different branch in $\type g$. In the former case, then $\term N$ is evaluated, with input $\type{!s}$ and output $\type{d}$. The overall term inputs $\type{!r}$ and outputs $\type{g ++ p.d}$, where the merge $\type{++}$ requires that whenever $\type g$ and $\type d$ have a common branch $j$, the associated stack types are the same. For the special case where $p=0$, i.e.\ $\term M$ chooses $i$ with probability zero, $\type{p.d}$ is the empty sub-distribution, and the rule becomes as follows.
\[
\infer{\term{G |- M;i->N : !r => g}}{\term{G |- M: !r => g} & (i\notin\floor{\type g})}
\]

	\item
The \emph{expand} rule implements a key principle of stack calculi, \emph{stack expansion}: if $\term{M}$ evaluates successfully with a stack $\type{!s}$, then if there are further inputs $\type{!r}$ on the stack, these will be returned unchanged. Recall that input type vectors are reversed, so that in the return type $\type{!s\,!r => !r\,d}$ the stack $\type{!r}$ is below $\type{!s}$ and below each stack in $\type d$. 

	\item
The \emph{iterate} rule requires that the input types $\type{!s}$ of $\term M$ match the output types on branch $i$ exactly, and that the probability $p$ of choosing $i$ is less than one. As with \emph{sequence} there is a special case where $p=0$, where the rule becomes as follows.
\[
\infer{\term{G |- M^i: !s => d}}{\term{G |- M: !s => d} & (i\notin\floor{\type d})}
\]

	\item
The \emph{sample} rule reflects that $\term{?a.M}$ behaves like an abstraction which may receive one of two values, $\term{\top : \e=>\e.1\top}$ or $\term{\bot : \e => \e.1\bot}$, which it binds to the variable $\term a$. The term $\term M$ is typed separately for each possibility, with the same input stack $\type{!r}$. The return type $\type{(/12.g)++(/12.d)}$ again requires that $\type g$ and $\type d$ have the same stack type for any choice they share.
\end{itemize}

\begin{example}
The probabilistic sum $\term{M+N}=\term{?a.a ; \top->M ; \bot->N }$ is typed as follows.
\[
\infer{ \term{|- ?a.a ; \top -> M ; \bot -> N : !t ~=>~ /12.g ~++~ /12.d} }{
 \infer{ \term{|- ?a.a ; \top -> M : !t ~=>~ /12.g ~*~ !t./12\bot } }{
  \infer{ \term{|- ?a.a : !t ~=>~ !t./12\top ~*~ !t./12\bot} }{
   \infer{ \term{|- ?a.a : \e ~=>~ \e./12\top ~*~ \e./12\bot} }{
    \infer{ \term{ a: \e => \e.1\top |- a: \e => \e.1\top} }{}
    &&
    \infer{ \term{ a: \e => \e.1\bot |- a: \e => \e.1\bot} }{}
  }} &
  \term{|- M : !t => g}
 } &
 \term{|- N : !t => d}
}
\]
Using this as an admissible typing rule, our running example is typed as follows.
\[
\infer{ \term{|- (\heads + (\tails + \,i))^i : \e ~=>~ \e./23\heads ~*~ \e./13\tails} }{
 \infer{ \term{|- \heads + (\tails + \,i) : \e ~=>~ \e./12\heads ~*~ \e./14\tails ~*~ \e./14i} }{
  \infer{ \term{|- \heads : \e ~=>~\e.1\heads}}{}
  &&
  \infer{ \term{|- \tails + \,i : \e ~=>~ \e./12\tails ~*~ \e./12i } }{
   \infer{ \term{|- \tails : \e ~=>~\e.1\tails}}{}
   &&
   \infer{ \term{|- i : \e ~=>~\e.1i}}{}
}}}
\]
\end{example}

\begin{remark}
\label{rem:probabilistic types embed}
The type system is a conservative extension of those of our previous work on probabilistic termination in~\cite{Antonelli-DalLago-Pistone-2022} and~\cite{Heijltjes-Majury-2025}. Using the notation of the latter,  these feature probabilistic simple types as simple types with the rationals between $0$ and $1$ as the set of base types, as below left. They embed as below right, where $\star$ is again the single available choice label.
\[
	 \typ{s,t} ~\coloneqq~ p~\mid~\typ{s\to t}
\qquad\qquad
	 \typ{t_1 \to \dots \to t_n \to p} \quad=\quad \type{t_1 \dots t_n => \e.p\star}
\]
The \emph{sequential probabilistic types} of~\cite{Heijltjes-Majury-2025} are the restriction of the present system to the default choice $\star$.
\[
	\typ{!s} \overset{p}{\Rightarrow} \typ{!t} \quad=\quad \type{!s => !t.p\star}
\]
\end{remark}

\begin{remark}
An element $\type{!t.pi}$ of a distribution type gives a lower bound $p$ on the probability of terminating with branch $i$. Since our types are a probabilistic form of simple types, which cannot type all normalizing $\lambda$-terms, the bound is necessarily not exact. Consider a term $\term M$ which is terminating but not typeable, for example $\term{[<y>.[y].*].<x>.[x].x}$, a variant of $\Omega=(\lambda x.xx)(\lambda y.yy)$ that replaces the function $y$ with $\term*$. Then $\term M$ terminates with probability $1$, but can only be typed by the zero-rule as $\term{M:\e=>0}$, giving a sum probability of zero. The zero-rule allows such terms to be combined with others, and express for instance that $\term{M+*}$ terminates with probability at least $\frac 12$. To obtain exact bounds, one would need a type system that can type all terminating $\lambda$-terms, such as intersection types.
\end{remark}

\section{Almost-Sure Termination}
\label{sec:AST}

The intended meaning of the type system is that if $\term{|- M: \e => d}$ (i.e.\ $\term M$ is closed and requires no inputs to run on the machine) then the probability of termination for $\term M$ is at least $\pr{\type d}$. We will establish this using a reducibility argument in the style of Tait~\cite{Tait-1967}, called \emph{runnability} as it uses the machine rather than reduction. The runnability predicate directly interprets the meaning of types: a runnable term of type $\term{M:!s => d}$, given a stack $S$ of runnable terms for the types in $\type{!s}$, will probabilistically evaluate to a runnable return family $\DT$ for the type $\type{d}$, with the key property that $\pr{\DT}\geq\pr{\type d}$.

\begin{definition}
The set $\RUN{t}$ of \emph{runnable terms} for a type $\type t$ is defined as the set of closed terms
\[
	\RUN{!s => d} = \{ \term M ~\mid~ \forall S \in\RUN{!s}.~\exists\DT\in\RUN{d}.~\Eval SM\DT~\}
\]
where the runnable sets for stack types $\type{!t}$ and distribution types $\type{d}$ are as follows.
\[
    \RUN{t_1..t_n} = \{ \e\,\term{M_1}\dots \term{M_n} \mid \term{M_i}\in\RUN{t_i} \}
\]
\[
	\RUN{\{(!t_i,p_i)_i\}_{i\in I}} = \{ \DT \mid \forall i\in I.~\stk{\DT\only i}\subseteq\RUN{!t_i}~\wedge~\pr{\DT\only i}\geq p_i~\}
\]
\end{definition}

\noindent
To work with open terms, a \emph{substitution map} $\mu$ is a finite function from variables to terms, applied to a term as $\term{\mu M}$ which replaces every free variable $x$ in $\term M$ that is in the domain of $\mu$ with the term $\term{\mu(x)}$. For a context $\Gamma=\term{x_1:t_1},\dots,\term{x_n:t_n}$ the set $\SRUN{G}$ is a set of substitution maps $\mu$ that assigns each $x_i$ a runnable term $\term M\in\RUN{t_i}$, as follows.
\[
	\SRUN{G}~=~\{ \mu \mid \forall\term{x:t}\in\Gamma.~\term{\mu(x)}\in\RUN{t}\}
\]
We are now ready to prove the \emph{runnability lemma}, that typed terms are runnable.

\begin{lemma}
\label{lem:run}
If $\term{G |- M:t}$ and $\mu\in\SRUN{G}$ then $\term{\mu M}\in\RUN{t}$.
\end{lemma}

\begin{proof}
\ifLongVersion
See Appendix~\ref{A:runnability}.
\else
See the extended version~\cite{DalLago-Heijltjes-2027-long}.
\fi
\end{proof}

The runnability lemma directly gives the main theorem of typed almost-sure termination.

\begin{theorem}
\label{thm:termination}
If $\term{|- M: \e => d}$ then $\term M$ terminates with probability at least $\pr{\type{d}}$.
\end{theorem}

\begin{proof}
Let $\type{d}=\type{\{(!t_i,p_i)_i\}_{i\in I}}$. By Lemma~\ref{lem:run} we have $\Eval\e M\DT$ with $\DT\in\RUN{d}$, which means that $\pr{\DT\only i}\geq p_i$ for every $i\in I$, so that $\pr\DT \geq \pr{\type d}$.
\end{proof}

\section{Ground Terms and Finite Markov Chains}
\label{sec:ground terms}

We have shown that the calculus introduced in this paper enjoys all the expected properties, in particular guaranteeing almost-sure termination. Several natural questions concerning the expressive power of the calculus remain open. Notably, it is of interest to investigate the behavior of the calculus under restrictions to low-order types, or when stronger termination properties than AST, such as \emph{positive} almost-sure termination (PAST), can be proved. In this section we resume our discussion of ground terms from the introduction.
\[
	\text{\emph{Ground terms:}}\qquad \term{M,N}~\coloneqq~\term i~\mid~\term{M;i->N}~\mid~\term{M^i}~\mid~\term{?a.a}
\]
We will show the following properties: one, typing of ground terms is decidable; two, ground terms readily encode as finite Markov chains; and three, finite Markov chains may be encoded as ground terms. These then have the following consequences. The last two properties imply that ground terms are equivalent to finite Markov chains. Next, since finite Markov chains are PAST (if they are AST), also typed ground terms are PAST. Finally, combining the first and last property, for the encoding of a Markov chain as a ground term, type inference calculates the probability of terminating with each final state.

As in the introduction, we may restrict types for ground terms to sub-distributions over choice labels, and abbreviate them by omitting input and return types.
\[
	\term{M : p_1i_1 * .. * p_ni_n}\quad\text{\emph{abbreviates}}\quad\term{M : \e => \e.p_1i_1 * .. * \e.p_ni_n}
\]
Since ground terms are closed, the typing rules may further omit contexts. Thus restricted, we recover the typing rules from the introduction (with new notation).
\[
       \infer{\term{i: 1i}}{}
\qquad \infer{\term{M;i->N : g ++ (p . d)}}{\term{M: g * pi} & \term{N: d}}
\qquad \infer{\term{M^i : /1{1-p} . d}}{\term{M: d * pi} & (p\neq 1)}
\qquad \infer{\term{?a.a : /12\top * /12\bot}}{}
\]

\begin{proposition}
\label{prop:ground type inference}
Typing for ground terms is decidable.
\end{proposition}

\begin{proof}
Immediate by induction on the term.
\end{proof}

We will define Markov chains as probabilistic automata, as follows. We write $\DD X$ for the set of probability distributions over a set $X$, and write a distribution over $X=\{x_1,x_2,\dots\}$ as $\{p_1x_1,p_2x_2,\dots\}$.

\begin{definition}
A \emph{Markov chain} $C=(S,s,F,\delta)$ is given by a set of \emph{states} $S$, a \emph{starting state} $s\in S$, a set of \emph{final states} $F\subseteq S$, and a stochastic \emph{transition} function $\delta:S\less F\to\DD{S}$ from non-final states to states.
\end{definition}

The translation $\markov M$ from ground terms to Markov chains will take $\term M$ to a chain
\[
	\markov{M}~=~(S_M,~s_M,~F_M,~\delta_M)
\]
where internal states $S_M\less F_M$ represent subformulas and final states represent choices not captured by a later case construct. We consider chains modulo renaming of internal states, and when combining the translations of two terms we assume their internal states to be disjoint, but final states may overlap. To connect two chains, we use the notation $s\{t/i\}$ to rename a state, and the notation $\delta\{s/i\}$ which replaces all transitions to a (final) state $i$ with transitions to $s$.
\[
	\delta\{s/i\}(r)~=~\{~p(t\{s/i\}) \mid pt \in \delta(r)~\}
\qquad\qquad
	\begin{array}{r@{}l} i\{t/i\} &{}=t \\ s\{t/i\} &{}=s \quad (s\neq i) \end{array}
\]
The translation $\markov-$ is then as follows.
\[
\begin{aligned}
	\markov{i}      &= (\{s,i\},~s,~\{i\},~s\mapsto\D{i.1})
\\	\markov{M;i->N} &= (S_M\less\{i\}\cup S_N,~s_M,~F_M\less\{i\}\cup F_N,~\delta_M\{s_N/i\}\cup\delta_N)
\\  \markov{M^i}    &= (S_M\less\{i\},~s_M,~F_M\less\{i\},~\delta_M\{s_M/i\})
\\  \markov{?a.a}   &= (\{s,\top,\bot\},~s,~\{\top,\bot\},~s\mapsto\D{/12\top,/12\bot})
\end{aligned}
\]

\begin{example}
For our running example term $\term M$ the translation is as follows.
\[
\begin{array}{r@{~}l}
          \term{M}~= & \term{(?a.a ; \top->\heads ; \bot->(?b.b ; \top->\tails ; \bot -> i))^i}
\\[8pt] \markov{M}~= & (~\{a,b,r,s,u,\heads,\tails\}~,~r~,~\{\heads,\tails\}~,~
\\[3pt]              & ~~a\mapsto\D{/12r,/12b}~~r\mapsto\D{1\heads}
\\[3pt]              & ~~b\mapsto\D{/12s,/12u}~~s\mapsto\D{1\tails}
\\[3pt]              & ~~u\mapsto\D{1r}~)
\end{array}
\qquad\qquad
	\vc{
\begin{tikzpicture}[thick,x=20pt,y=24pt,outer sep=0pt,inner sep=2pt]
	\node (r) at (1,2) {$a$};
	\node (s) at (2,1) {$b$};
	\node (u) at (0,1) {$r$};
	\node (v) at (1,0) {$s$};
	\node (w) at (3,0) {$u$};
	\node (h) at (0,0) {$\termcolor\heads$};
	\node (t) at (1,-1) {$\termcolor\tails$};

	\begin{scope}[->,>=stealth]
	
	\draw (1,2.7)--(r);
	\draw (r) --node[pos=0.6,anchor=south east]{$\scriptstyle{\frac12}$} (u);
	\draw (r) --node[pos=0.6,anchor=south west]{$\scriptstyle{\frac12}$} (s);
	\draw (s) --node[pos=0.6,anchor=south east]{$\scriptstyle{\frac12}$} (v);
	\draw (s) --node[pos=0.6,anchor=south west]{$\scriptstyle{\frac12}$} (w);
	\draw (u)--(h);
	\draw (v)--(t);
	
	\draw[rounded corners] (w)--(3.7,0.7)--(1.7,2.7)--(r);
	
	\end{scope}
\end{tikzpicture}}
\]
\end{example}

In the other direction, we give an interpretation $\sem C$ of a finite Markov chain as a ground term. The translation operates as follows, using states as choice labels. Given $C$ with starting state $s$ and transition $s\mapsto\{p_1t_1,\dots,p_nt_n\}\in\delta$, we remove this transition and consider the sub-chains $C(t_i)$ for $i\leq n$ with new starting state $t_i$, in which $s$ becomes a final state. These are translated inductively to $\term{M_i}$. Then $\sem C$ is given by a term $\term{M^s}$, iterating on $s$, where $\term M$ is a probabilistic sum that chooses $\term{M_i}$ with probability $p_i$. We formalize this below.

For $C=(S,s,F,\delta)$ and $t\in S$ let $C(t)=(S,t,F\cup\{s\},\delta\less\{s\})$ where $\delta\less X$ is the transition function $\delta$ restricted to the domain $\mathrm{dom}(\delta)\less X$. Write $\term{p_1M_1+..+p_nM_n}$ for the term that chooses $\term{M_i}$ with probability $p_i$, given inductively below; recall that $\term{?a_p}$ represents an unfair coin toss, given either as primitive or encoded with $\term{?a}$ and iteration.
\[
	\term{1M}~=~\term{M} 
\qquad
    \term{pM+q_1N_1+..+q_nN_n}~=~\term{?a_p.a ; \top -> M; \bot -> (//{q_1}{1-p}N_1 + .. + //{q_n}{1-p}N_n) } 
\]
Then the interpretation of a Markov chain $C=(S,s,F,\delta)$ as a term is as follows.
\[
	\sem C~=~
	\left\{\begin{array}{ll}
		\term{s} & \text{if}~s~\in~F
	\\	\termcolor{(p_1\sem{C(t_1)}\term{+..+}p_n\sem{C(t_n)})^s} & \text{where}~\delta(s)=\D{p_1t_1,..,p_nt_n}
	\end{array}\right.
\]

\begin{example}
\label{ex:markov}
Consider the following Markov chain $C$, where each arrow has probability $\frac13$, and its typed interpretation $\sem C$, writing $\term{M+N+P}$ for the term $\term{//13M+//13N+//13P}$. A fully worked version is in Appendix~\ref{A:markov}.
\[
\vcenter{\hbox{
\begin{tikzpicture}[inner sep=2pt,outer sep=0pt,thick]
	\node[circle,draw] (a) at ( 90:0.6) {$a$} ;
	\node[circle,draw] (b) at (210:0.6) {$b$} ;
	\node[circle,draw] (c) at (330:0.6) {$c$} ;
	\node (i) at ( 90:1.4) {$i$} ;
	\node (j) at (210:1.4) {$j$} ;
	\node (k) at (330:1.4) {$k$} ;
	\begin{scope}[->,>=stealth]
	\draw (a.240) to[bend right] (b.60) ;
	\draw (b.0)   to[bend right] (c.180) ;
	\draw (c.120) to[bend right] (a.300) ;
	\draw (a.330) to[bend left]  (c.90) ;
	\draw (b.90)  to[bend left]  (a.210) ;
	\draw (c.210) to[bend left]  (b.330) ;
	\draw (a)--(i);
	\draw (b)--(j);
	\draw (c)--(k);
	\draw (120:1.2)--(a);
	\end{scope}
\end{tikzpicture}}}
\qquad
\begin{array}{@{}r@{}l@{}}
	    C ~=~{}& (~\{a,b,c,i,j,k\},~a,~\{i,j,k\},~
\\[3pt]	       & ~~ a \mapsto \D{/13b,/13c,/13i}~~b \mapsto \D{/13a,/13c,/13j}~~c \mapsto \D{/13a,/13b,/13k}~)
\\[10pt]
   \sem C ~=~{}& \term{((a+(a+b+k)^c+j)^b + (a+(a+c+j)^b+k)^c+i)^a}: \type{/12i * /14j * /14k}
\end{array}
\]
\end{example}

\noindent
The translations $\markov-$ and $\sem-$ above demonstrate the following.

\begin{proposition}
\label{prop:ground is markov}
Ground terms of the \PFMC\ are equivalent to finite Markov chains.
\end{proposition}

\noindent
Markov chains are PAST if they are AST, i.e.\ if any path may eventually reach a final state. For typed ground terms, which are AST, we then have the following.

\begin{proposition}
\label{prop:PAST}
Typed ground terms with sum probability $1$ of the \PFMCT\ are PAST.
\end{proposition}

\noindent
Combining type inference for ground terms (Proposition~\ref{prop:ground type inference}) and the translation $\sem-$, we have the following result.

\begin{proposition}
\label{prop:markov return}
Encoding and typing a finite Markov chain computes its return distribution: if ${\termcolor\sem{C}}:\type{p_1`s_1*..*p_n`s_n}$ then $C$ terminates in the state $s_i$ with probability $p_i$.
\end{proposition}

\begin{proof}
By construction.
\end{proof}

\section{Moving past PAST}
\label{sec:NAST}

Further exploring the expressive power of the calculus, in this section we demonstrate that typed terms in general might be \emph{null AST} (NAST), i.e.\ AST, but not PAST. We will give an example term with infinite expected running time, despite terminating with probability one. Our example relies crucially on higher-order functions.
Consider the following term, for a given $\term M$, using the default choice $\star$ as imperative \emph{skip} and its case as sequential composition, $\term{M;N}=\term{M;\star ->N}$. 
\[
	\term{[M].(\,(<f>.[f;f].i)\, + \,\star)^i }
\]
With probability $\fraction12$ it composes the term on the stack with itself and iterates, or it terminates. Evaluating this term on the machine gives the following return family, for the sequence of terms $\term{M_n}$ given by $\term{M_0}=\term M$ and $\term{M_{n+1}}=\term{M_n;M_n}$.
\[
	\{(\term{M_n},\star)_{\top^n\bot}\mid n\in\N\}~=~\{\,(\term M,\star)_{\bot}~,~(\term{M;M},\star)_{\top\bot}~,~(\term{(M;M);(M;M)},\star)_{\top\top\bot}~,~\dots~\}
\]
We may then evaluate the term $\term{M_n}$ left on the stack by post-composing with $\term{<g>.g}$, which pops and runs it, possibly using further base values further down on the stack. If $\term M$ terminates successfully with $\star$ in $m$ steps, $\term{M_n}$ takes $m\times 2^n$ steps. For example, assuming a natural numbers type $\type{\N}$ with primitives zero $\term{0:\N}$ and addition $\term{`+:\N\,\N => \N.1\star}$, we let $\term M$ be the successor function $\term{[1].`+: \N => \N.1\star}$. The term below left then gives the return family below right, taking $m\times 2^n$ machine steps (for some small constant $m$) to compute the number $\term{2^n:\N}$ from zero and successor. Since the probability of returning $\term{2^n}$ is $\fraction 1{2^{n+1}}$, the expected evaluation time on the machine is infinite: the term is not PAST.
\[
	\term{[0].[[1].`+].(\,(<f>.[f;f].i)\, + \,\star)^i ; <g>.g ~:~\e => \N . 1\star }
\qquad\qquad
	\{ (\term{2^n},\star)_{\top^n\bot}\mid n\in\N \}
\]
The term is typed: for any $\type{!t}$, if $\term M$ is typed $\term{M:!t => !t.1\star}$ then we have $\term{M^n : !t => !t.1\star}$, such as in the case above where $\type{!t}=\type\N$. For the loop we then have the following.
\[	
	\vc{\infer{\term{|- ((<f>.[f;f].i)\,+\,\star)^i: s => s.1\star}}{\term{|- (<f>.[f;f].i)\,+\,\star: s => s./12i ~*~ s./12\star }}} \qquad \text{where} \quad \type{s}~=~\type{!t => !t.1\star}
\]
The term $\term{<g>.g : (!t => !t.1\star)\,!t => !t.1\star}$ takes $\term{M^n : !t => !t.1\star}$ off the stack and evaluates it with an appropriate remaining stack of type $\type{!t}$. The existence of a typed term that is not PAST then demonstrates the following.

\begin{proposition}
\label{prop:NAST}
The typed probabilistic FMC is not positive almost-surely terminating.
\end{proposition}

\section{Limitations and Prospects}
\label{sec:limits}

The key to our type system is that it enforces a fixed probability of exiting or continuing for each iteration of a loop. Naturally, this limits expressiveness, as it prevents the typing of iteration with growing or shrinking probabilities, which might nevertheless be AST. Here, we explore how to express such algorithms in the untyped calculus, examine why they cannot be typed, and consider the prospects of lifting this restriction. 

Consider an unfair coin toss $\term{M:\top *_p \bot}$, using a (further) abbreviated notation $\type{\top *_p \bot}$ for the type $\type{\e=>\e.p\top * \e.(1`-p)\bot}$. We may construct terms with modified probabilities for example as follows.
\[
\begin{aligned}
 	\term{[a]_q.a\,;\,\top -> M} &~:~\type{\top *_{pq} \bot}
&&&	\term{M;\top ->M}            &~:~\type{\top *_{p^2} \bot}
\end{aligned}
\]
To alter the probability of exiting a loop during iteration, we make essential use of the higher-order nature of the calculus: we store the intended loop body on the stack, and modify it during iteration. Using the second construction above:
\[
	\term{[M]\,.\,( <x>.(x\,;\,\bot -> [x;\top -> x].*)^\star}
\]
The term starts with $\term M$ on the stack; pops and evaluates it; on $\top$, it exits the loop; on $\bot$, it pushes $\term{M;\top ->M}$ to the stack and iterates. The probability to exit the loop squares at each iteration, from $p$ to $p^2$, $p^4$, $p^8$, $p^{16}$, etc. The probability of non-termination is $\prod_{n\in\mathbb N}(1-p^{2^n})$, which is non-zero e.g.\ for $p=\frac12$: the term is not AST.

The reason why the term cannot be typed is subtle: it is because the loop body changes the probability on the stack. The body itself can be typed, for any probability $q$:
\[
	\term{<x>.(x\,;\,\bot -> [x;\top -> x].*)~:~(\top *_q \bot) => \e.q\top * (\top *_{q^2} \bot).(1`-q)\star }
\]
But to type the loop requires the same input and output types on the stack, while $\type{\top *_q \bot}$ and $\type{\top *_{q^2} \bot}$ are distinct.

This observation explains how the type system enforces fixed probabilities during iteration: the probabilities of all inputs---the stack---need to remain fixed. This remains a main limitation of the type system, as it also prevents typing similar terms to the above example that are AST, for example one that increases (rather than decreases) the probability of exiting the loop. Finally, the above considerations indicate how the type system might be extended to accommodate iteration with changing probability: by introducing variables for probabilities, and arithmetic functions over them, into the type system, approaching a form of dependent types.


\section{Conclusion}

We have demonstrated a type system for almost-sure termination in a probabilistic $\lambda$-calculus with iteration, the \PFMCT. As a direct probabilistic generalization of the FMC and typed iteration, we believe this to be a natural account of AST.
essive power beyond positive AST.

In future work, we intend to further analyse the expressive power of the calculus and fragments of it. At present, being an extension of simple types with fixed probabilities of iteration, it is clear that the expressive power of the typed calculus for real algorithms is limited. Addressing this, we expect that the type system will generalize with polymorphism, dependent types, and intersection types along standard lines. A dependently typed generalisation would be akin to Hoare logic, and might facilitate reasoning about probabilistic termination in theorem provers and give natural ways to include termination evidence in typed imperative programming languages.


\bibstyle{plainurl}
\bibliography{FMC-AST}

\newpage

\appendix

\section{On function composition and recursion}
\label{A:recursion}

As the introduction set out, typed almost-sure termination arises naturally by replacing the sum in the standard typed iterator with a probabilistic sum. The probabilistic FMC integrates iteration into the $\lambda$-calculus, giving AST in a higher-order setting. The $\lambda$-calculus is however more commonly associated with \emph{recursion} than iteration. A pertinent question is then whether typed AST is also possible for recursion. Here, we consider the obstacles to typing probabilistic termination in a probabilistic $\lambda$-calculus without branching and iteration. We will use the type system for the probabilistic event $\lambda$-calculus, given by arrow types over probabilities $p\in\mathbb Q\cap[0,1]$.
\[
	\type{s,t}~\coloneqq~\type p~\mid~\type{s\to t}
\]

First, we compare function composition,
\[
	g\circ f~=~\lambda x.\,g\,(f\,x)
\]
against sequencing $\term{f;g}$. The particular problem is to type self-composition, $f\circ f$ (the Church numeral two). Suppose we have $f:\type{1\to/12}$, i.e.\ $f$ returns with probability $\frac12$, so that we need $f\circ f:\type{1\to/14}$. For input $x:\type 1$ we get $f\,x:\type{/12}$, as expected, but to apply $f$ again we need to modify its type to $f:\type{/12\to/14}$. This may be approached in several ways.
\begin{itemize}
	\item
Using \emph{intersection types}: we assign $f$ a collection of multiple types $f:\{\type{1\to/12},\type{/12\to/14}\}$. This is used in~\cite{Antonelli-DalLago-Pistone-2022}, among others.
	\item
Using \emph{subtyping}: we may introduce a subtyping relation that coerces $f:\type{1\to/12}$ to $f:\type{/12\to/14}$ by multiplying all probabilities by a fixed factor, in this case $\frac12$. This is a rather coarse approach, since it ignores the actual flow of computation in a term, and only works as long as the type system does not multiply any probabilities.
	\item
Using \emph{polymorphic types}: with probabilistic type variables we may assign the type $f:\type{p\to/12p}$, which instantiates to $f:\type{1\to/12}$ and $f:\type{/12\to/14}$.
\end{itemize}

While these are all reasonable ideas, they are not as simple and natural as our current approach using sequential composition and iteration, which has the intrinsic ability to multiply the probabilities of sequential computations:
\[
	\term{f:t=>t./12\star} \qquad \text{and}\qquad \term{f;f : t => t./14\star}
\]

Turning to recursion, consider the following two recursive terms $f$ and $g$, where $1:\type 1$ is a terminating constant and $\Omega=(\lambda x.xx)(\lambda x.xx):\type 0$ diverges.
\[
	f~=~1 \oplus (\Omega\oplus f) \qquad\qquad g~=~g\oplus g
\]
Analogous to our running example from the introduction, $f$ should return with probability $\frac23$, and indeed we are able to assign that as its type:
\[
	f:\type{/23} ~\vdash~ 1 \oplus (\Omega\oplus f): \type{/23}
\]
However, we are also able to type $g$, which diverges, with any probability $p$:
\[
	g:\type{p} ~\vdash~ g\oplus g: \type{p}
\]
The key problem is that with recursion, termination happens through \emph{weakening}, i.e.\ discarding the recursive function call. In $f$, the terminating branch $1$ discards the recursive call, while in $g$ neither branch does, but this difference is invisible to the type system. Whether a subterm will be discarded during computation is an undecidable property typically tracked using intersection types, and not usually registered in a system of simple types.

A solution, therefore, would need to register the multiple branches of a computation, and which variables are weakened or not. For example, with \emph{polymorphism} we might have the following types,
\[
	f:\type{p} ~\vdash~ 1\oplus(\Omega\oplus f): \type{/12\oplus /14p}
\]
and solve $p=\frac 12+\frac14p$ for $p=\frac23$ in the rule for fixed points.

This is an interesting direction to explore, and we may do so in future. However, unlike our approach for iteration, it requires moving beyond simple types, and is beyond the scope of the present work.


\section{Running the machine}
\label{A:run}

In Figure~\ref{fig:run} we give a successful run of the machine for our running example term, unfolded to its primitives, and the trace $\bot\,\bot\,\top$.
\[
	\term{M}~=~\term{(\heads + (\tails +\,i))^i}~=~\term{(?a.a ; \top->\heads ; \bot->(?b.b ; \top->\tails ; \bot -> i))^i}
\]

\begin{figure}
\[
\scalebox{0.975}{$
\begin{array}{@{(\,}r@{\,,\,}l@{\,,\,}l@{\,,\,}r@{\,)}}
             \bot\,\bot\,\top & \e &         \term{(?a.a ; \top->\heads ; \bot->(?b.b ; \top->\tails ; \bot -> i))^i} & \e
\\\hline     \bot\,\bot\,\top & \e & \phantom(\term{?a.a ; \top->\heads ; \bot->(?b.b ; \top->\tails ; \bot -> i)} & \term{(i -> M)}
\\\hline     \bot\,\bot\,\top & \e & \phantom(\term{?a.a ; \top->\heads} & \term{(\bot->(?b.b ; \top->\tails ; \bot -> i))~(i -> M)}
\\\hline     \bot\,\bot\,\top & \e & \phantom(\term{?a.a} & \term{(\top->\heads)~(\bot->(?b.b ; \top->\tails ; \bot -> i))~(i -> M)}
\\\hline           \bot\,\top & \e & \phantom(\term{\bot} & \term{(\top->\heads)~(\bot->(?b.b ; \top->\tails ; \bot -> i))~(i -> M)}
\\\hline           \bot\,\top & \e & \phantom(\term{\bot} &                \term{(\bot->(?b.b ; \top->\tails ; \bot -> i))~(i -> M)}
\\\hline           \bot\,\top & \e & \phantom(\term{?b.b ; \top->\tails ; \bot -> i} &                               \term{(i -> M)}
\\\hline           \bot\,\top & \e & \phantom(\term{?b.b ; \top->\tails} &                               \term{(\bot -> i)~(i -> M)}
\\\hline           \bot\,\top & \e & \phantom(\term{?b.b} &                               \term{(\top->\tails)~(\bot -> i)~(i -> M)}
\\\hline                 \top & \e & \phantom(\term{\bot} &                               \term{(\top->\tails)~(\bot -> i)~(i -> M)}
\\\hline                 \top & \e & \phantom(\term{\bot} &                                              \term{(\bot -> i)~(i -> M)}
\\\hline                 \top & \e & \phantom(\term{i}    &                                                          \term{(i -> M)}
\\\hline                 \top & \e & \term{(?a.a ; \top->\heads ; \bot->(?b.b ; \top->\tails ; \bot -> i))^i} & \e
\\\hline                 \top & \e & \phantom(\term{?a.a ; \top->\heads ; \bot->(?b.b ; \top->\tails ; \bot -> i)} & \term{(i -> M)}
\\\hline                 \top & \e & \phantom(\term{?a.a ; \top->\heads} & \term{(\bot->(?b.b ; \top->\tails ; \bot -> i))~(i -> M)}
\\\hline                 \top & \e & \phantom(\term{?a.a} & \term{(\top->\heads)~(\bot->(?b.b ; \top->\tails ; \bot -> i))~(i -> M)}
\\\hline                   \e & \e & \phantom(\term{\top} & \term{(\top->\heads)~(\bot->(?b.b ; \top->\tails ; \bot -> i))~(i -> M)}
\\\hline                   \e & \e & \phantom(\term{\heads} &              \term{(\bot->(?b.b ; \top->\tails ; \bot -> i))~(i -> M)}
\\\hline                   \e & \e & \phantom(\term{\heads} &                                                        \term{(i -> M)}
\\\hline                   \e & \e & \phantom(\term{\heads} & \e
\end{array}$}
\]
\caption{A complete run of the machine}
\label{fig:run}
\end{figure}


\section{Encoding Markov chains}
\label{A:markov}

We elaborate on Example~\ref{ex:markov}, encoding the following Markov chain $C$, where each arrow has probability $\frac13$.
\[
\vcenter{\hbox{
\begin{tikzpicture}[inner sep=2pt,outer sep=0pt,thick]
	\node[circle,draw] (a) at ( 90:1) {$a$} ;
	\node[circle,draw] (b) at (210:1) {$b$} ;
	\node[circle,draw] (c) at (330:1) {$c$} ;
	\node (i) at ( 90:2) {$i$} ;
	\node (j) at (210:2) {$j$} ;
	\node (k) at (330:2) {$k$} ;
	\begin{scope}[->,>=stealth]
	\draw (a.240) to[bend right] (b.60) ;
	\draw (b.0)   to[bend right] (c.180) ;
	\draw (c.120) to[bend right] (a.300) ;
	\draw (a.330) to[bend left]  (c.90) ;
	\draw (b.90)  to[bend left]  (a.210) ;
	\draw (c.210) to[bend left]  (b.330) ;
	\draw (a)--(i);
	\draw (b)--(j);
	\draw (c)--(k);
	\draw (120:2)--(a);
	\end{scope}
\end{tikzpicture}}}
\qquad
\begin{array}{l@{}l@{}l}
	    S      &{}~=~{}& \{a,b,c,i,j,k\}
\\[3pt]	s      &{}~=~{}& a
\\[3pt]	F      &{}~=~{}& \{i,j,k\}
\\[3pt]	\delta &{}~=~{}& a \mapsto \D{~/13b~,~/13c~,~/13i~}
\\[3pt]               && b \mapsto \D{~/13a~,~/13c~,~/13j~}
\\[3pt]               && c \mapsto \D{~/13a~,~/13b~,~/13k~}
\end{array}
\]
The interpretation, in stages, is as follows, again writing $\term{M+N+P}$ for $\term{//13M+//13N+//13P}$.
\[
\begin{aligned}
	\sem C &~=~ \term{({\sem{C(b)}}+{\sem{C(c)}}+i)^a}
\\	       &~=~ \term{((a+{\sem{C(b)(c)}}+j)^b + (a+{\sem{C(c)(b)}}+k)^c+i)^a}
\\         &~=~ \term{((a+(a+b+k)^c+j)^b + (a+(a+c+j)^b+k)^c+i)^a}
\end{aligned}
\]
The type is built up as follows.
\[
\begin{aligned}
	    \term{(a+b+k)^c}         &: \type{/13a * /13b * /13k}
\\[5pt]	\term{ a+(a+b+k)^c+j}~~~ &: \type{/49a * /19b * /19k + /13j}
\\[5pt] \term{(a+(a+b+k)^c+j)^b} &: \type{/12a * /18k * /38j}
\\[5pt] \term{(a+(a+c+j)^b+k)^c} &: \type{/12a * /18j * /38k}
\\[5pt] \term{ (a+(a+b+k)^c+j)^b + (a+(a+c+j)^b+k)^c+i}~~~ &: \type{/13a * /16k * /16j * /13i}
\\[5pt] \term{((a+(a+b+k)^c+j)^b + (a+(a+c+j)^b+k)^c+i)^a} &: \type{/14k * /14j * /12i}
\end{aligned}
\]
The interpretation may be illustrated as follows, where the solid edges represent the tree-shape of the probabilistic sums, while the dashed edges represent the loops.
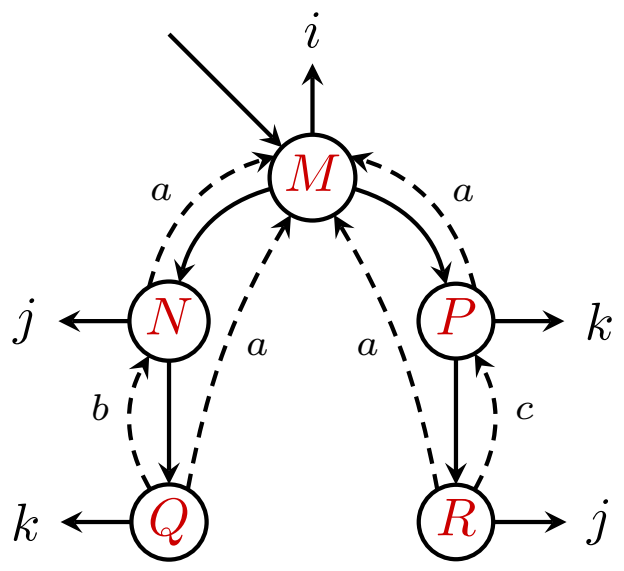
\[
\vc{
\begin{tikzpicture}[x=27pt,y=-27pt,inner sep=0.5pt,outer sep=0pt,thick]
	\node[circle,draw] (a) at (1,0) {$\term M$};
	\node[circle,draw] (b) at (0,1) {$\term N$};
	\node[circle,draw] (c) at (2,1) {$\term P$};
	\node[circle,draw] (d) at (0,2.4) {$\term Q$};
	\node[circle,draw] (e) at (2,2.4) {$\term R$};
	\node (i) at (1,-1)   {$\vphantom(i$};
	\node (j) at (-1,1)   {$~j~$};
	\node (k) at ( 3,1)   {$~k~$};
	\node (l) at (-1,2.4) {$~k~$};
	\node (m) at ( 3,2.4) {$~j~$};
	\begin{scope}[->,>=stealth]
	\draw (0,-1)--(a);
	\draw (a) to[bend right] (b);
	\draw (a) to[bend left]  (c);
	\draw (a) -- (i);
	\draw (b) -- (d);
	\draw (c) -- (e);
	\draw (b) -- (j);
	\draw (c) -- (k);
	\draw (d) -- (l);
	\draw (e) -- (m);
	\begin{scope}[dash pattern=on 3pt off 2pt]
	\draw (b.120) to[bend left]     node[pos=0.5,anchor=south east]{$\scriptstyle a~$} (a.150);
	\draw (c.60)  to[bend right]    node[pos=0.5,anchor=south west]{$~\scriptstyle a$} (a.30);
	\draw (d.120) to[bend left]     node[pos=0.5,anchor=south east]{$\scriptstyle b~$} (b.240);
	\draw (e.60)  to[bend right]    node[pos=0.5,anchor=south west]{$~\scriptstyle c$} (c.300);
	\draw (d.60)  to[bend left=10]  node[pos=0.5,anchor=west]      {$~\scriptstyle a$} (a.240);
	\draw (e.120) to[bend right=10] node[pos=0.5,anchor=east]      {$\scriptstyle a~$} (a.300);
	\end{scope}
	\end{scope}
\end{tikzpicture}
}
\qquad
\begin{array}{l@{}l}
        \term M &{}~=~\term{(N + P + i)^a}
\\[3pt] \term N &{}~=~\term{(a + Q + j)^b}
\\[3pt] \term P &{}~=~\term{(a + R + k)^c}
\\[3pt] \term Q &{}~=~\term{(a+b+k)^c}~=~\term{a+b+k}
\\[3pt] \term R &{}~=~\term{(a+c+j)^b}~=~\term{a+c+j}
\end{array}
\]
As the illustration suggests, the formalism may express any probabilistic tree with back-pointers. It may also express directed acyclic graphs, which yields a smaller interpretation of the Markov chain above, though still exponential, in the following term.
\[
	\term{ (b+c+i ; b -> a+c+j ; c -> (a+b+k ; b -> a+c+j)^c )^a }
\]
The types work out the same way.
\[
\begin{aligned}
	    \term{ b+c+i }~~~             &: \type{/13b * /13c * /13i}
\\[5pt] \term{ b+c+i ; b -> a+c+j}~~~ &: \type{/19a * /49c * /13i * /19j}
\\[5pt] \term{(a+b+k ; b -> a+c+j)^c} &: \type{/12a * /18j * /38k}
\\[5pt] \term{ b+c+i ; b -> a+c+j ; c -> (a+b+k ; b -> a+c+j)^c}     &: \type{/13a * /13i * /16j * /16k}
\\[5pt] \term{(b+c+i ; b -> a+c+j ; c -> (a+b+k ; b -> a+c+j)^c )^a} &: \type{/12i * /14j * /14k}
\end{aligned}
\]


\ifLongVersion

\section{Proving the runnability lemma}
\label{A:runnability}

Here we prove Lemma~\ref{lem:run}, the runnability lemma. Recall the statement of the Lemma:

\medskip

\noindent
\textbf{Runnability lemma:}\quad\textit{
If $\term{G |- M:t}$ and $\mu\in\SRUN{G}$ then $\term{\mu M}\in\RUN{t}$.
}

\begin{proof}
By induction on the typing derivation.
\begin{itemize}

	\item \emph{Variable:} 
\[
\infer{\term{G, x:t |- x:t}}{}
\]
By the definition of $\SRUN{G}$ we have $\term{\mu x}\in\RUN t$.

	\item \emph{Choice:}
\[
\infer{\term{G |- i: \e => \e.1i}}{}
\]
Note that $\RUN{\e}=\{\e\}$ and $\term{\mu i}=\term i$ for any $\mu$. We have the evaluation $\eval\e\e i\e i$ for the empty trace and the empty stack. Then for the frontier $\{\e\}$ we have $\Eval \e i{\{(\e,i)_\e\}}$, with the singleton return family of $\e$ and $i$ for the empty trace. Since $p(\e)=1$ it follows that $\{(\e,i)_\e\}\in\RUN{\e.1i}$.

	\item \emph{Zero}
\[
\infer{\term{G |- M : \e => 0}}{}
\]
For any $\mu$ we have $\Eval \e{\mu M}\varnothing$. Vacuously, $\varnothing\in\RUN{0}$, since membership is defined by the quantification $\forall i\in\floor{\type 0}$ while the latter is empty.

	\item \emph{Expand:}
\[
\infer{\term{G |- M: !s\,!r => !r\,d}}{\term{G |- M: !s => d}}
\]
Let $R\in\RUN{!r}$, $S\in\RUN{!s}$, and $\mu\in\SRUN{G}$. The inductive hypothesis gives the evaluation $\Eval SM\DT$. For each $(T,i)_t\in\DT$ we have the evaluation below left. By induction on the evaluation relation $\evalarrow$ we get the evaluation below right.
\[
\eval tS{\mu M}Ti
\qquad\qquad
\eval t{RS}{\mu M}{RT}i
\]
This gives $\Eval{RS}{\mu M}{R\,\DT}$. Since $R\in\RUN{!r}$ and $\stk{\DT\only i}\subseteq\RUN{!t_i}$, where $\type{!t_i}$ is the type for $i$ in $\type d$, we have $\stk{R\,\DT\only i} \subseteq \RUN{!r\,!t_i}$. Since also $\pr{R\,\DT\only i}=\pr{\DT\only i}\geq p_i$ we have $R\,\DT\in\RUN{!r\,d}$. 

	\item \emph{Push:} 
\[
\infer{\term{G |- [N].M : !s => d}}{\term{G |- N:r} && \term{G |- M : r\,!s => d}}
\]
Let $S\in\RUN{!s}$ and $\mu\in\SRUN{G}$. By induction we have $\term{\mu N}\in\RUN{r}$ and $\term{\mu M}\in\RUN{r\,!s=>d}$, and for the latter the evaluation $\Eval {S\,\term{\mu N}}{\mu M}\DT$. Then for every $(T,i)_t\in \DT$ we have the evaluation below left, from which we construct the derivation below right.
\[
	\eval t{S\,\term{\mu N}}{\mu M}Ti
\qquad\qquad
	\vc{\infer[\Rpush]{ \eval tS{\mu([N].M)} Ti }{ \eval t{S\,\term{\mu N}}{\mu M} Ti } }
\]
Then $\Eval S{\mu([N].M)}\DT$ and $\term{\mu([N].M)}\in\RUN{!s => d}$.

	\item \emph{Pop:}
\[
\infer{\term{G |- <x>.M : r\,!s => d}}{\term{G , x:r |- M: !s => d}}
\]
Let $S\in\RUN{!s}$, $\term N\in\RUN r$, and $\mu\in\SRUN{G}$. Let $\mu'=\mu\{\term N/x\}$, the substitution map that sends $x$ to $\term N$ and every other variable $y$ to $\mu y$, so that $\mu'\in\SRUN{G,x:r}$. By induction we have $\Eval S{\mu'M}\DT$, given by the evaluation below for every $(T,i)_t\in\DT$, from which we construct the evaluation below right.
\[
	\eval tS{\mu'M} Ti
\qquad\qquad
	\vc{ \infer[\Rpop]{ \eval t{S\,\term{N}}{\mu(<x>.M)}Ti } {\eval tS{\mu'M}Ti} } 
\]
Then $\Eval{S\,\term N}{\mu(<x>.M)}\DT$ and $\term{\mu(<x>.M)}\in\RUN{r\,!s=>d}$.

	\item \emph{Sample:}
\[
\infer{\term{G |- ?a.M : !r => (/12 . g) ++ (/12 . d)}}{\term{G , a: \e => \e.1\top |- M: !r => g} && \term{G , a: \e => \e.1\bot |- M: !r => d}}
\]
Let $R\in\RUN{R}$ and $\mu\in\SRUN{G}$. Define $\mu_\top=\mu\{\term\top/a\}$ and $\mu_\bot=\mu\{\term\bot/a\}$ so that $\mu_\top\in\SRUN{G , a: \e => \e.1\top}$ and $\mu_\bot\in\SRUN{G , a: \e => \e.1\bot}$. By induction we have return families $\DS$ and $\DT$ where $\Eval R{\mu_\top M}\DS$ and $\Eval R{\mu_\bot M}\DT$, given by the following evaluations for every $(S,i)_s\in\DS$ and $(T,j)_t\in\DT$.
\[
\eval sR{\mu_\top M}Si \qquad\qquad  
\eval tR{\mu_\bot M}Tj
\]
For these we have the following derivations.
\[
\infer[\Rtop]{ \eval{\top\,s}R{\mu(?a.M)}Si }{ \eval sR{\mu_\top M}Si } \qquad
\infer[\Rbot]{ \eval{\bot\,t}R{\mu(?a.M)}Tj }{ \eval tR{\mu_\bot M}Tj }
\]
Then for $\DU=(\top\pfx\DS)\cup(\bot\pfx\DT)$ we have $\Eval S{\mu(?a.M)}\DU$. It remains to show that $\DU\in\RUN{/12.g ++ /12.d}$. For any indexed pair in $\type{/12.g ++ /12.d}$, by the definition of $(\type{++})$ we have three cases.
\begin{itemize}
	\item $\type{(!t,/12p+/12q))_i}$ where $i\in\floor{\type g}\cap\floor{\type d}$, $\type{(!t.p)_i}\in\type g$ and $\type{(!t.q)_i}\in\type d$. Then since $\stk{\DS\only i}\subseteq\RUN{!t}$ and $\stk{\DT\only i}\subseteq\RUN{!t}$ we have $\stk{\DU\only i}\subseteq\RUN{!t}$, and since $\pr{\DS\only i}\geq p$ and $\pr{\DT\only i}\geq q$ we have $\pr{\DU\only i}=\pr{\top\pfx\DS\only i}+\pr{\bot\pfx\DT\only i}\geq \frac 12p+\frac 12q$.
	\item $\type{(!t,/12p)_i}$ in $\type{/12.g-{\floor\delta}}$. Since $\stk{\DS\only i}\subseteq\RUN{!t}$ we have $\stk{\DU\only i}\subseteq\RUN{!t}$, and since $\pr{\DS\only i}\geq p$ we have $\pr{\DU\only i}=\pr{\top\pfx\DS\only i}\geq\frac 12p$.
	\item The case for $\type{/12.d-{\floor\gamma}}$ is analogous to the previous.
\end{itemize}

	\item \emph{Sequence:} 
\[
\infer{\term{G |- M;i->N : !r => g ++ p.d}}
      {\term{G |- M: !r => g * !s.pi} && 
       \term{G |- N: !s => d}}
\]
Let $R\in\RUN{!r}$ and $\mu\in\SRUN{G}$. By induction we have $\term{\mu M}\in\RUN{!r => g * !s.pi}$ and $\Eval R{\mu M}\DS$. Let this be given by an evaluation for each $(S,j)_s\in\DS$ of the form below left. For each $(S,j)_s$ where $j=i$ we have $S\in\RUN{!s}$, and by induction on $\term{N}$ a return family $\DT_s$ such that $\Eval S{\mu N}{\DT_s}$. Let this be given by the evaluation below right for each $(T,k)_t\in\DT_s$.
\[
	\eval sR{\mu M}Sj
\qquad
	\eval tS{\mu N}Tk
\] 
For each $(S,j)_s$, if $j\neq i$  we construct the derivation below left, and if $j=i$ then for each $(T,k)_t\in\DT_s$ we construct that below right. 
\[
	\infer[\Rrej]{ \eval sR{\mu(M;i->N)}Sj }{ \eval sR{\mu M}Sj }
\qquad\qquad
	\infer[\Rsel]{ \eval{st}R{\mu(M;i->N)}Tk }{ \eval sR{\mu M}Sj && \eval tS{\mu N}Tk }
\]
Then we have $\Eval R{\mu(M;i->N)}\DT$ where $\DT$ is given as follows.
\[
	\DT~=~(\DS\less i) \cup \bigcup_{s\in\floor{\DS\only i}} s\pfx\DT_s
\]
It remains to show that $\DT\in\RUN{g ++ p.d}$.

\begin{itemize}
	\item
First, for the return types, for all $j\in\floor{\type g}\cup\floor{\type d}$ we need that $\stk{\DT\only j}\subseteq\RUN{!t}$ where $\type{!t}$ is the type for $j$ in $\type g$ and $\type d$ (the definition of $(\type{++})$ requires that they agree). By definition any indexed pair in $\DT$ is either $(S,j)_s\in\DS\less i$ or $(T,k)_t\in\DT_s$. The former case follows since $\DS\in\RUN{g}$ and so $\stk{\DS\only j}\subseteq\RUN{!t}$ and $S\in\subseteq\RUN{!t}$. The latter case follows since $\DT_s\in\RUN{d}$, so that $\stk{\DS\only j}\subseteq\RUN{!t}$ and $T\in\RUN{!t}$.
	\item
Second, for the probabilities, for all $j\in\floor{\type g}\cup\floor{\type d}$ it must be shown that $\pr{\DT\only j}\geq p_j + pq_j$ where $p_j$ is the probability of $j$ in $\type g$ and $q_j$ that in $\type d$. We compute as follows.
\[
\begin{aligned}
\pr{\DT\only j} 
    &~=~    \pr{((\DS\less i)~\cup~\bigcup_{s\,\in\,\floor{\DS\only i}} s\pfx\DT_s)\only j}
\\  &~=~    \pr{(\DS\less i)\only j} + \sum_{s\in\floor{\DS\only i}} (\pr{s}\times\pr{\DT_s\only j})
\\  &~\geq~ p_j + \sum_{s\in\floor{\DS\only i}} (\pr{s}\times q_j)
\\  &~\geq~ p_j + pq_j
\end{aligned}
\]
For the third step (the first inequality), since $\DS\less i\in\RUN{g}$ we have $\pr{\DS\only j}\geq p_j$ for all $j\in\floor{\type g}$. For the last step, since $\DT_s\in\RUN{d}$ for all $s$ we have $\pr{\DT_s\only j}\geq q_j$ for all $j\in\floor{\type d}$. Then since $\DS\only i\in\RUN{!s.pi}$ we have $\pr{\DS\only i}\geq p$.
\end{itemize}

It follows that $\DT\in\RUN{g ++ p.d}$ and hence $\term{\mu(M;i->N)}\in\RUN{!r => g ++ p.d}$.

	\item \emph{Iterate:}
\[
\infer{\term{G |- M^i: !s => /1{1-p} . d}}{\term{G |- M: !s => d + !s.pi} && (p\neq 1)}
\]
Let $S\in\RUN{!s}$ and $\mu\in\SRUN{G}$. We construct a series of return families $\DT_n$ for all $n\in\N^+$ by induction on $n$, and show the following.
\[
	\DT_n\in\RUN{(q_n.d) + !s.p^ni} \quad\text{where}\quad q_n=1+p+p^2+p^3+\dots+p^{n-1}
\]

\begin{itemize}
\item
The case $\DT_1$ is given by induction on $\term M$ by $\Eval S{\mu M}{\DT_1}$. 

\item
For the case $\DT_{n+1}$, for each $(R,i)_r\in\DT_n$, by induction on $\term M$ let $\Eval R{\mu M}{\DT_r}$ and define $\DT_{n+1}$ as follows.
\[
	\DT_{n+1}~=~(\DT_n\less i) \cup \bigcup_{r\in\floor{\DT_n\only i}} r\pfx\DT_r
\]
By induction we have $\DT_n\in\RUN{q_n.d + !s.p^ni}$. By similar reasoning to the previous \emph{sequencing} case we have the following.
\[
	\DT_{n+1}\in\RUN{q_n.d + p^n.(d + !s.pi)}~=~\RUN{q_{n+1}.d + !s.p^{n+1}i}
\]
\end{itemize}

Observe that by definition $\DT_n\less i~\subseteq~\DT_{n+1}\less i$. Then we define $\DT$ as the following supremum.
\[
	\DT~=~\sup_{n\in\N^+}\DT_n\less i
\]
It remains to show that $\Eval S{\mu(M^i)}\DT$ and $\DT\in\RUN{/1{1-p}.d}$.

\begin{itemize}
	\item
If $(T,j)_t\in\DT$, then by definition $(T,j)_t\in\DT_n\less i$ for some $n$. Assume this is the least such $n$. Then by induction, going backwards from $n$, we give a sequence of stacks $S_m$ and traces $s_m$ and $t_m$ for all $m<n$ such that $S_0=S$, $s_0=\e$, $s_m\,t_m=t$, and $\eval {t_m}{S_m}{\mu(M^i)} Tj$, and $(S_m,i)_{s_m}\in\DT_m$ for $m\neq 0$.

\begin{itemize}
	\item
For the case $m=n-1$, since $(T,j)_t\in\DT_n$, by definition it is either in $\DT_{n-1}\less i$, which is ruled out by assumption, or it is given by an evaluation $\Eval R{\mu M}{\DT_r}$ with $R=S$ if $n=1$, and otherwise $(R,i)_s\in\DT_{n-1}$ and $t=sr$, with the evaluation below left. Let $S_{n-1}=R$, $s_{n-1}=s$, and $t_{n-1}=r$. Since $i\neq j$, this gives the derivation below right.
\[
\eval rR{\mu M} Tj \qquad \infer[\Rexit]{ \eval {t_{n-1}}{S_{n-1}}{\mu(M^i)} Tj }{ \eval {t_{n-1}}{S_{n-1}}{\mu M} Tj }
\]

	\item
For the cases $0<m<n$, by induction let $(S_m,i)_{s_m}\in\DT_m$ with $t=s_mt_m$ and the evaluation below left. By the definition of $\DT_m$, $(S_m,i)_{s_m}$ is given by an evaluation $\Eval R{\mu M}{\DT_r}$ with $R=S$ if $m=1$ and otherwise $(R,i)_s\in\DT_{m-1}$ with $s_m=sr$, giving the evaluation below right.
\[
	\eval{t_m}{S_m}{\mu(M^i)}Tj
\qquad\qquad
	\eval rR{\mu M}{S_m}i
\] 
To show the statement for $m-1$, let $S_{m-1}=R$, $s_{m-1}=s$, and $t_m=r\,t_{m-1}$, and construct the derivation below.
\[
	\infer[\Rloop]{ \eval {t_{m-1}}{S_{m-1}}{\mu(M^i)} Tj }
	{ \eval {s_{m-1}}{S_{m-1}}{\mu M} {S_m}i && \eval {t_m}{S_m}{\mu(M^i)} Tj }
\]
\end{itemize}
It follows that for every $(T,j)_t\in\DT$ there is an evaluation $\eval tS{\mu(M^i)}{T_t}{j_t}$, and hence $\Eval S{\mu(M^i)}\DT$.

	\item
For all $n\in\N^+$ we have $\DT_n\in\RUN{q_n.d + !s.p^ni}$ and hence $\DT_n\less i\in\RUN{q_n.d}$. Since each $T$ in $\stk{\DT\only j}$ is in some $\stk{\DT\only j}$ we have $T\in\RUN{!t}$ where $\type{!t}$ is the stack for $j$ in $\type d$. Next, for $j\neq i$ we have $\pr{\DT_n\only j} \geq q_np_j$ where $q_n=1+p+p^2+\dots +p^{n-1}$ and $p_j$ is the probability for $j$ in $\type{d}$. This gives the following. 
\[
	\pr{\DT\only j} \geq (1+p+p^2+p^3\dots)\times p_j ~=~ \frac1{1-p}\,p_j 
\]
\end{itemize}

It follows that $\DT\in\RUN{/1{1-p}.d}$ and hence $\term{\mu(M^i)}\in\RUN{!s => /1{1-p}.d}$.
\end{itemize}
\end{proof}

\fi

\end{document}